\documentclass[preprint,superscriptaddress,showpacs,preprintnumbers,amsmath,amssymb,aps, pre]{revtex4-1}

\usepackage[utf8]{inputenc}
\usepackage{subfigure}
\usepackage{graphicx}
\usepackage{dcolumn}
\usepackage{bm}
\usepackage{enumitem}
\usepackage{xcolor}
\usepackage{csquotes}

\begin{document}

\title{How to detect Wada Basins}

\author{Alexandre~Wagemakers}
\email{Corresponding author: alexandre.wagemakers@urjc.es}
\affiliation{Nonlinear Dynamics, Chaos and Complex Systems Group, Departamento de  F\'isica, Universidad Rey Juan Carlos\\ Tulip\'an s/n, 28933 M\'ostoles, Madrid, Spain}

\author{Alvar~Daza}
\affiliation{Nonlinear Dynamics, Chaos and Complex Systems Group, Departamento de  F\'isica, Universidad Rey Juan Carlos\\ Tulip\'an s/n, 28933 M\'ostoles, Madrid, Spain}
\affiliation {Department of Physics, Harvard University, Cambridge, Massachusetts 02138, USA  }

\author{Miguel A.F. Sanju\'{a}n}
\affiliation{Nonlinear Dynamics, Chaos and Complex Systems Group, Departamento de  F\'isica, Universidad Rey Juan Carlos\\ Tulip\'an s/n, 28933 M\'ostoles, Madrid, Spain}
\affiliation{Department of Applied Informatics, Kaunas University of Technology, Studentu 50-415, Kaunas LT-51368, Lithuania}
\date{\today}

\begin{abstract}

We present a review of the different techniques available to study a special kind of fractal basins of attraction known as Wada basins, which have the intriguing property of having a single boundary separating three or more basins. We expose several approaches to identify this topological property that rely on different, but not exclusive, definitions of the Wada property.
\end{abstract}

\maketitle

\section{\label{sec:Introduction}Introduction}

The origin of Wada basins dates back to 1917, when Kunizo Yoneyama published a work on topology where he described how to divide a region of the plane in three or more connected sets sharing a common boundary \cite{yoneyama_theory_1917}. He attributed the authorship of the original procedure to his advisor Takeo Wada, and since then these intricate topological constructions were called Wada lakes. At first, the intriguing properties of Wada lakes were studied within a topological context \cite{hocking1988topology}. For example, the Polish topologist Kazimierz Kuratowski showed that if a boundary separates at the same time three or more connected regions in the plane, then the boundary must be an indecomposable continuum \cite{kuratowski_sur_1924, sanjuan1997indecomposable}. Years later, Wada lakes were studied by James Yorke and collaborators under the perspective of dynamical systems \cite{kennedy_basins_1991, nusse_wada_1996}. They analyzed the set of initial conditions leading to a particular attractor, called the basins of attraction, in a forced damped pendulum. The authors demonstrated numerically that for a particular set of parameters, the forced damped pendulum presents three basins of attraction sharing the same boundary, that is, they are Wada basins. The Nusse-Yorke condition to assert the Wada property in \cite{nusse_wada_1996} was based on the computation of the unstable manifold of a saddle point, which intersected all the three basins. This is how an apparently inconceivable geometry arose in such a simple system as the forced damped pendulum. The cumbersome structure of the Wada basins implies a particular kind of unpredictability \cite{daza_basin_2016}, since a small perturbation in the initial conditions lying on a Wada boundary may lead the trajectory to any of the system's attractors. Since the pioneering works of Yorke and collaborators \cite{kennedy_basins_1991, nusse_saddle-node_1995, nusse_wada_1996, nusse_fractal_2000}, the Wada property has been found in many different cases: chaotic scattering \cite{poon_wada_1996,hansen2018statistical}, Hamiltonian systems \cite{aguirre_wada_2001}, fluid dynamics \cite{toroczkai_wada_1997}, interaction between waves \cite{coninck2007basins}, delayed systems \cite{daza_wada_2017}, black hole shadows  \cite{BHs}, etc.

In most of these works, the authors used the Nusse-Yorke condition mentioned earlier. However, Daza et al. \cite{daza_testing_2015, daza_ascertaining_2018, wagemakers2020saddle} have recently proposed three new methods to test for the Wada property. Each one relies on a different perspective of Wada basins and, consequently, they extend our understanding of this property. Also, these three algorithms can reduce considerably the computational efforts and enable the identification of the Wada property in a wider variety of systems and situations. The main goal of this paper is to review the essential properties of each of these three methods, providing a comparison of their main features. The information is organized as follows. First, we describe the Nusse-Yorke computational method that tracks an unstable manifold of a unstable periodic orbit. In Sec.~\ref{sec_grid} we describe the grid approach, a numerical test based on the successive refining of the grid. Section~\ref{sec:MergingMethod} is devoted to the merging method, a quick graphical test to detect Wada basins. Last but not least, the saddle-straddle method to identify Wada basins using the chaotic saddle is presented in sec.~\ref{Sec.straddle}. A good description of some of the invariant sets involved in these methods can be found in \cite{tel_2006}. Furthermore, all the methods are illustrated through several paradigmatic examples. Finally, we conclude comparing their main advantages and drawbacks.

\section{\label{sec_NY_method} Crossing three basins: the Nusse-Yorke method}

We should start with a historically important method that has been the only one available for many years. It exposes an interesting connection between the Wada property and the presence of unstable periodic orbits in the observable phase space. To assure that the basin is Wada, the following condition must be fulfilled:

\begin{displayquote}
{\it Condition 1: If $P$ is an unstable periodic orbit accessible from a basin $B_1$, its unstable manifold must intersect every basin.}
\end{displayquote}

\begin{figure}
\begin{center}
\includegraphics[width=0.6\textwidth]{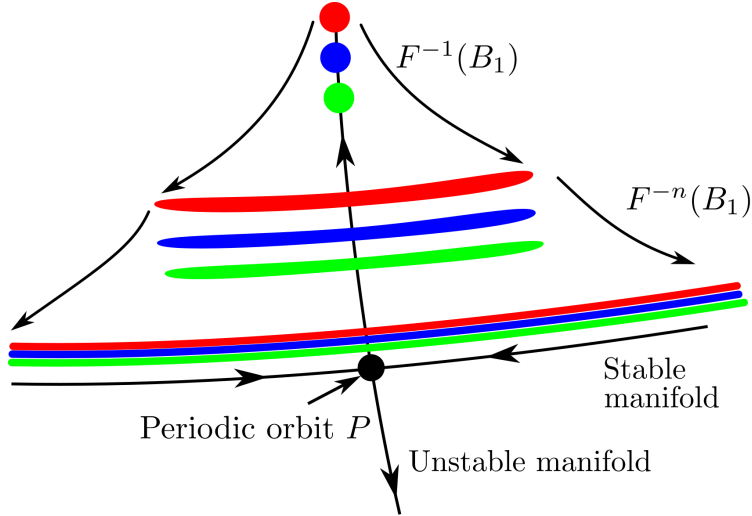}
\end{center}
\caption{\label{fig_NY_1} {\bf Sketch of the condition 1 of the Nusse-Yorke method}. The small disks represent areas of the basins $B_1$, $B_2$ and $B_3$. The unstable manifold of the unstable periodic orbit $P$ intersects the three basins. The preimages of the disks are stretched exponentially and asymptotically approach the stable manifold, which ultimately is the Wada boundary.}
\end{figure}

It is possible to understand why this condition is necessary with a simple picture of a two-dimensional phase space with three basins $B_1$, $B_2$ and $B_3$. Suppose an unstable periodic orbit in the phase plane labeled $P$ in Fig.~\ref{fig_NY_1} and its unstable and stable manifold. The unstable manifold of $P$ intersects the three basins represented by three small disks of different colors. If we compute the preimages $F^{-1}(B_j)$ of these small sets under the action of the dynamical system as time goes backward, we observe a stretching of the sets along the stable manifold and a contraction along the unstable manifold. As time goes backward, the preimages approach successively the stable manifold and become exponentially stretched. The repetition of the operation $n$ times leaves us with an image of a layered sequence of basins $B_1$, $B_2$ and $B_3$ each time closer to the stable manifold. In the limit, all points on the stable manifold of $P$ are arbitrarily close to the three basins, therefore, the stable manifold of $P$ is a Wada boundary.

Unfortunately, the condition 1 is a necessary but not sufficient condition to assure that the basin has the Wada property. The system may present other unstable periodic orbits which do not fulfill condition 1. In this case, we have only partially Wada basins. To assure that the basin is Wada one of the following conditions must be satisfied:

\begin{displayquote}
{\it Condition 2A: If there is a periodic orbit P that satisfies condition 1, the basin B satisfies the Wada property if the stable manifold of such saddle point is dense in the boundary of all basins.}
\end{displayquote}

\begin{displayquote}
{\it Condition 2B: If there is a periodic orbit P that satisfies condition 1, the basin B satisfies the Wada property if such saddle point is the only accessible orbit from basin B. In case that there is more than one accessible periodic orbit; every unstable manifold must intersect all basins.}
\end{displayquote}

\begin{displayquote}
{\it Condition 2C: If there is a periodic orbit P that satisfies condition 1, the basin B satisfies the Wada property if such saddle point generates a basin cell.}
\end{displayquote}

Condition 2A is extremely difficult to verify even in the simplest cases. The second condition requires to find all accessible periodic orbits. If the system presents more than one, the unstable manifold of each one must intersect all basins. The last condition 2C involves a structure called basin cell which is a trapping region formed by pieces of the stable and unstable manifolds of a boundary periodic orbit. If a basin cell is found, it means that there is only one accessible orbit.

\subsection{Description of the Nusse-Yorke method}

The following routine is an attempt to go through the verification of the conditions described earlier. It should be fit for ODEs, Hamiltonian and maps.

\begin{enumerate}
\item First, we must have a graphical description of the basins on a finite grid.
\item Find as many accessible periodic orbits as possible.
\item Plot the unstable manifolds of every accessible periodic orbit and verify that they intersect all basins.
\item One of the following conditions must be checked:
\begin{enumerate}
\item Verify the density of the stable manifold of the accessible orbit. This is a  possibility but we do not have the numerical tool to do this.
\item Verify that all accessible orbits have been found. To accomplish this task we sweep through the phase space in order to be sure that there are no elusive orbits hidden.
\item Construct a basin cell. Plot the stable and unstable manifolds of the accessible orbit, and construct a trapping region. It is important to remark that basin cells are only present in dissipative systems. The existence of a basin cell assures that there is only one accessible periodic orbit.
\end{enumerate}
\end{enumerate}

The steps (1), (2), (3), (4b) and (4c) can be executed with available numerical packages such as Dynamics \cite{nusse2012dynamics}. However, the search for the unstable periodic orbit and the computation of the unstable manifolds requires mastering the software and a detailed study of the dynamical system. It is unclear if these tasks could be fully automated, but certainly it would not be straightforward. As a matter of fact, most of the work on Wada basins prior to 2015 has been devoted exclusively to apply the Nusse-Yorke condition to a particular system with fixed parameters at a time \cite{poon_wada_1996, vandermeer_wada_2004, aguirre_unpredictable_2002, aguirre_wada_2001, zhang_wada_2013}, given the difficulty of the application of this method to each particular case.

\subsection{\label{sec_NY:examples}Examples}

\begin{figure}
\begin{center}
\subfigure[]{\includegraphics[width=0.49\textwidth,clip=no]{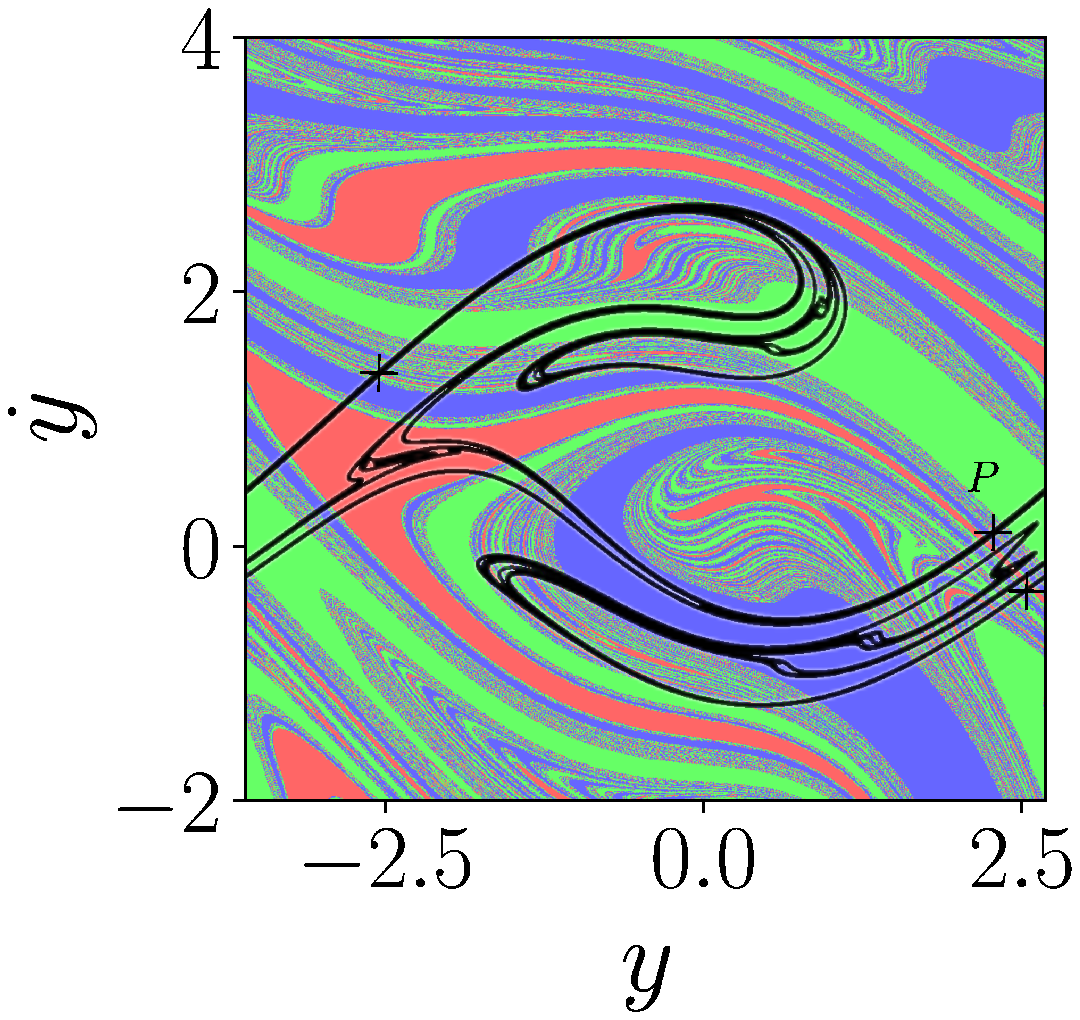}}
\subfigure[]{\includegraphics[width=0.49\textwidth]{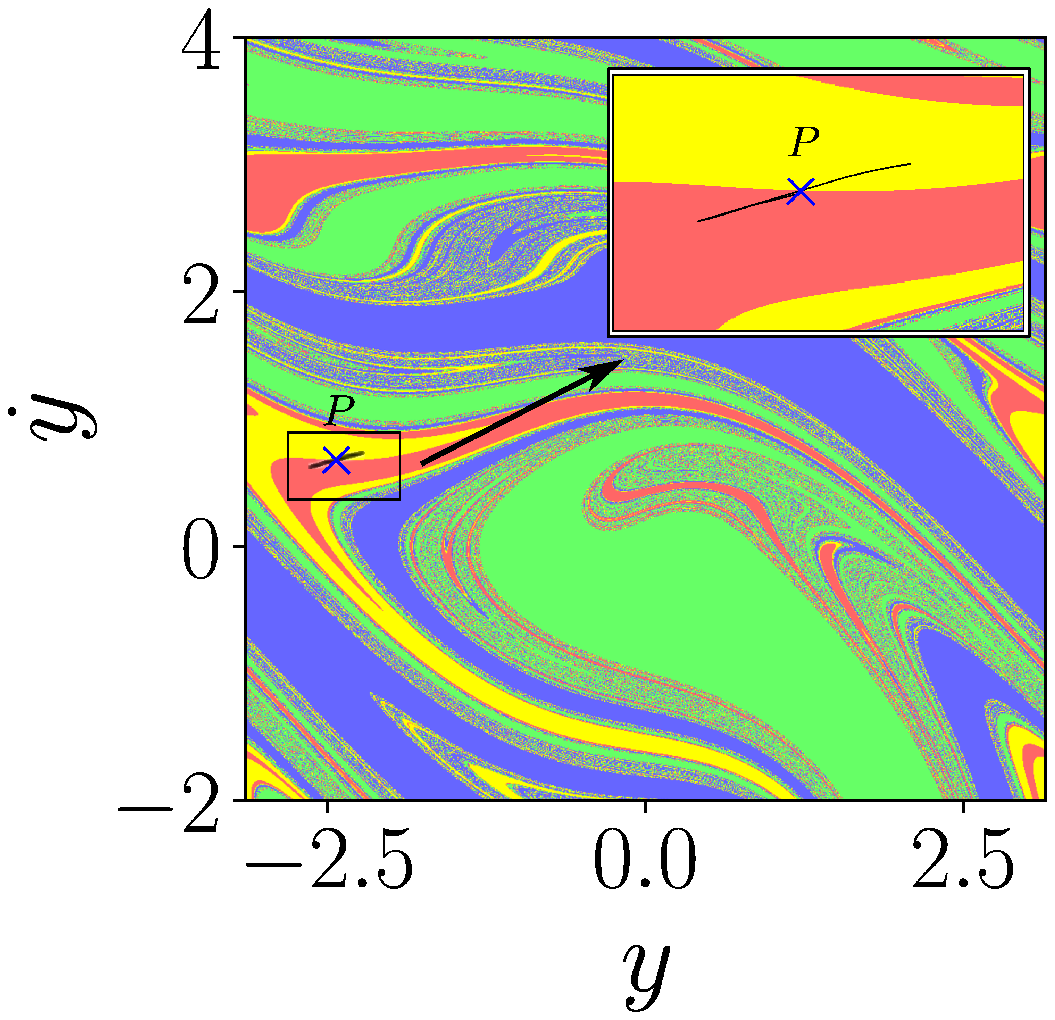}}
\end{center}
\caption{\label{fig_NY_2} {\bf Wada detection with the Nusse-Yorke method}
(a) Basins of attraction of the forced damped pendulum $\ddot{x}+0.2\dot{x}+\sin x=1.66\cos t$, with the unstable manifold of a period three orbit (crosses on the basins). The unstable manifold intersects the three basins. (b) Basins of attraction of the forced damped pendulum $\ddot{x}+0.2\dot{x}+\sin x=1.71\cos t$. There are four basins and we have found an accessible periodic orbit whose unstable manifold crosses only two basins. There is also a period-three periodic orbit similar to the case in (a). This basin is partially Wada.}
\end{figure}

For illustrative purposes, we present an application of the Nusse-Yorke method for the paradigmatic forced damped pendulum \cite{kennedy_basins_1991} in two different regimes. The first regime presents a fractal phase space with the Wada property, while the second example can only be classified as partially Wada. The forced damped pendulum is given by the equation
\begin{equation} \label{forced_dmp_eq}
\ddot{x}+0.2\dot{x}+\sin x=1.66\cos t,
\end{equation}
and it has three attractors that define three basins in its phase space $(x,\dot{x})$, depicted in Fig.~\ref{fig_NY_2}(a).

An exhaustive search of unstable periodic orbits shows a single period 3 orbit on the boundary of the basins. The unstable manifold is shown with black dots over the basin of attraction in Fig.~\ref{fig_NY_2}(a). It clearly intersects the three basins, we can therefore conclude that this basin is Wada, since condition (4c) has been positively checked.

However, increasing the forcing amplitude to $1.71$ leads to a different situation where the system does not exhibit the full Wada property. We have found two different unstable orbits on the boundary. One of these orbits has its unstable manifold plotted in Fig.~\ref{fig_NY_2}(b). This manifold only intersects two basins. The other unstable orbit on the boundary has its unstable manifold crossing the four basins (not shown). We can conclude that the basin is only partially Wada.

The numerical techniques used for these computations is the Quasi-Newton method with random initial seeds in the phase space to track the unstable orbits. The unstable manifolds are obtained iterating small segments very close to the saddle. These techniques are available in the numerical software Dynamics \cite{nusse2012dynamics}.

\section{\label{sec_grid}Down the scale: the grid approach}

Each of the numerical methods that we describe here rely upon a key observation on the properties of Wada basins that allows to establish a numerical test. But before starting, we proceed establish some conventional notation to describe the basins of attraction.

We will assume some simple and general hypothesis about the basins. First, we assume that there is a bounded region $\Omega$ containing $N_A\geq3$ disjoint regions $B_j$ where $j = 1, \cdots, N_A$. We also assume that there is a rectangular grid of $K$ boxes $P=\lbrace box_1,...,box_K\rbrace$ covering $\Omega$ whose interiors do not intersect each other. A typical grid would be $1000\times 1000$, thus $K\sim 10^6$.

We consider that it is possible to determine to which set $B_j$ belongs each point $x$ in $\Omega$. In other words, there is a function $C$ with $C(x) = j$ if $x \in B_j$ and $C(x)= 0$ if $x$ is in none of the sets $B_j$. If the sets are basins, the trajectory for each $x\in \Omega$ leads to an attractor labeled by $C(x)$. For any rectangular box denoted as $box$ we define $C(box) = C(x)$ where $x$ is the point at the center of the $box$. For convenience we will refer to this numerical value $C$ as the \textit{color} of the grid box. Of course other points in the same box might lead to different attractors.

For the method described hereafter, the important fact about Wada basin boundaries is the following:
\begin{displayquote}
{\it Given two different boxes $i$ and $j$ with different colors $C(i)\neq C(j)$, we will always find a third color between the two boxes if the boundary has the Wada property.}
\end{displayquote}
In the grid method, the algorithm looks for this third color by successive refinements of the basin until a stopping criterion has been met.

\begin{figure}
\begin{center}
\includegraphics[width=\textwidth]{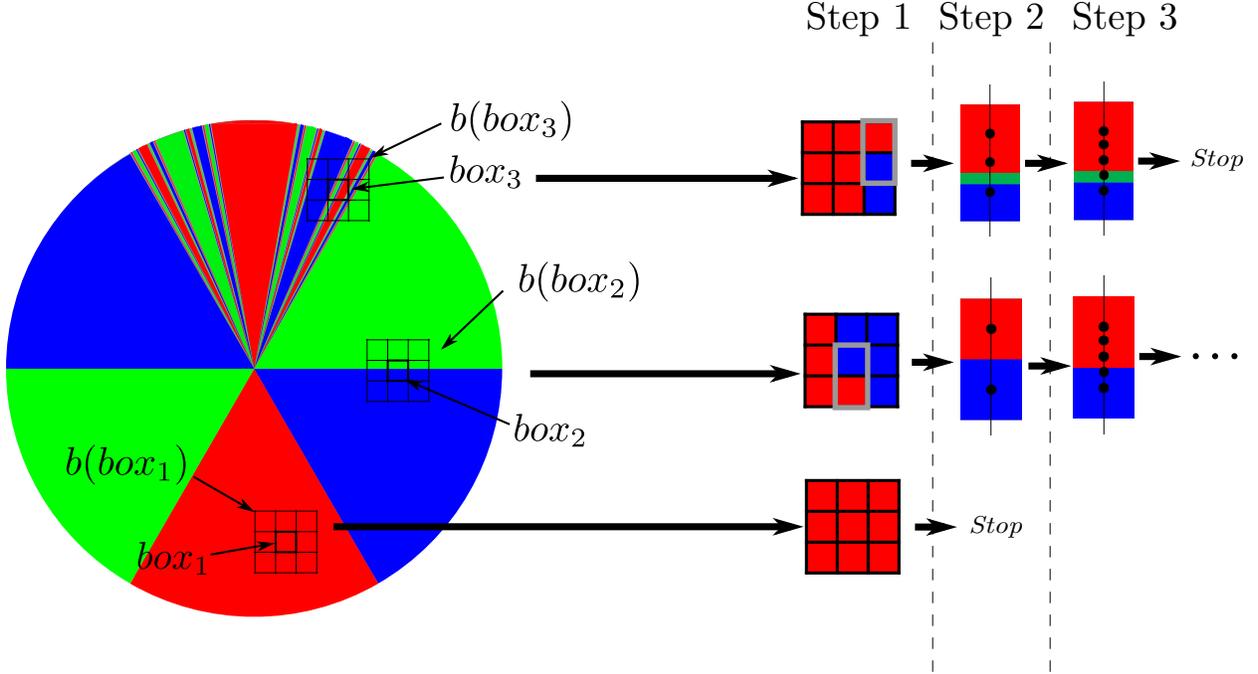}
\end{center}
\caption{\label{fig:method} {\bf Sketch of the grid method}. We set up a grid of boxes $box_j$ covering the whole disk. The center point of each box defines its color. In the first step, we see that $box_1$ belongs to the interior because its surrounding 8 boxes have the same color. On the other hand, $box_2$ and $box_3$ are in the boundary of two attractors, i.e., they are adjacent to boxes whose color is different. In the next step the algorithm classifies $box_2$ still in $G_2$ (boundary of two), while $box_3$ is now classified in $G_3$ (boundary of three). Ideally the process would keep on forever redefining the sets $G_1, G_2$ and $G_3$ at each step, though in practice we can impose some stopping condition. This plot constitutes an example of \textit{partially} Wada basins.  }
\end{figure}

\subsection{\label{sec_grid:method}Description of the grid method}

Before diving through the different scales looking for \textit{the third color}, we need to establish a reference grid, that will determine the accuracy of our algorithm. This reference grid is made of balls $b(box_j)$, which are the collection of grid boxes consisting of $box_j$ and all the grid boxes that have at least one point in common with $box_j$. Thus, in dimension two, $b(box_j)$ is a $3\times 3$ collection of boxes with $box_j$ being the central box. For each $box_j$, we determine the number of different (non-zero) colors in $b(box_j)$ and write $M(box_j)$ for that number.

In each $box_j$ with $M(box_j)\neq 1, N_A$, that is a box which is not in the interior nor in the Wada boundary, we accomplish the following procedure.
\begin{enumerate}
\item  We select the two closest boxes in $b(box_j)$ with different colors and trace a line segment between them. We compute the color of the middle point of the segment. In case that the color newly computed completes all colors inside $b(box_j)$, then $M(box_j)=N_A$ and the algorithm stops. Otherwise, we compute two new points in between the three previous ones.
\item In the second step, the color of four points interspersed with the previous five points is calculated. In the third step, we compute eight points interspersed with the previous nine and, in general, in the $nth$ step, $2^n$ new trajectories must be computed. This procedure keeps on until $M(box_j)=N_A$ or the number of calculated points in that segment reaches some maximum value previously set up. A major computational advantage of this method is that the refinement is made in a one-dimensional subspace (the segment linking the two points), no matter the dimension of $\Omega$.
\item  Next, we define $G_m$ to be the set of all the original grid boxes $box_j$ for which  $M(box_j) = m$. The number of elements in these sets $\# G_m$ is the output of the algorithm: they are the key to decide whether the basin is Wada or not.
\end{enumerate}

For $m=1$, all the boxes inside the ball $b(box_j)$ have the same color as they all lead to the same attractor, so $\# G_1$ is the number of boxes that are in the interior of a basin and is irrelevant for our purposes. The number $\# G_2$ is the number of boxes on the boundary of two basins, $\# G_3$ on the boundary of three basins and so on. To follow the evolution of these sets as the algorithm runs, we call $G_n^q$ the set $G_n$ at step $q$.

We say that the system is Wada if $ \lim\limits_{q \to \infty} \sum\limits_{m=2}^{{N_A}-1}\# G_m^q  =0$. This simply means that the grid boxes are either in the interior $G_1$ or in the Wada boundary $G_{N_A}$ after a sufficient number of steps $q$.

To illustrate the iterative process we represent an example of a partially Wada basin in Fig. \ref{fig:method}, and we compute the basin boundary for three grid boxes $box_1$, $box_2$, and $box_3$ on a regular rectangular grid. The first iteration for  $box_1$ shows that it belongs to the interior region $G^0_1$, since the eight boxes surrounding it have the same color. At this point, we can consider $box_1$ in $G^0_1$ without refining the partition. The second iteration, for $box_2$, lies in the boundary of two sets because two different colors are found in its ball $b(box_2)$. The subsequent iterations of the algorithm classify $box_2$ into $G_2$. A different situation arises for $box_3$. The first iteration classifies $box_3 \in G_2^0$, because only two colors are found in its ball. However, as far as we increase the resolution, $box_3$ turns out to be in the boundary of three basins $G_3^1$.

As previously stated, the basic idea underlying the whole process is that if three basins are Wada, then it is always possible to find a third color between the other two colors (similar reasoning can be done for Wada basins with more than three colors). Notice also that if a boundary separates two basins, then we will only see those two basins at every resolution.

In order to decide whether a system is Wada, not Wada, or presents an intermediate situation, we can count the number of boxes belonging to the boundary of $m$ different basins. For that purpose we define a useful parameter $W_m$ as,
\begin{equation}
W_m=\lim\limits_{q \to \infty} \frac{\# G_m^q }{\sum\limits_{j=2}^{N_A}\# G_j^q},
\end{equation}
where $m \in [2,N_A]$. This parameter $W_m \in [0,1]$ takes the value zero if the system has no grid boxes that are in the boundary separating $m$ basins, and it takes the value one if all the boxes in the boundary separate $m$ basins. Thus, if $W_{N_A}=1$ the system is said to be Wada. Partially Wada basins \cite{zhang2012unpredictability, zhang_wada_2013, zhang2014wada} occur when $0<W_m<1$ with $m \geq 3$, and this parameter provides a useful tool to classify them.

There is an alternative approach to the grid method developed in \cite{ziaukas2017fractal} and employed in \cite{lu2018control} which uses a fixed grid size $\varepsilon$ to compute the equivalent to the parameter $W_{N_A}$. There is no selective refinement of the grid to classify precisely the boxes. The result is an index $W$ called the Wada measure that is a number between 0 (smooth or partially Wada) and 1 (Wada). This is a less precise calculation but much faster as it does not check if the boxes have been correctly classified.

\subsection{\label{sec_grid:examples} Examples}

We present an application of the grid method with the forced damped pendulum in two different regimes presented in the sec. \ref{sec_NY:examples}.

\begin{figure}
\begin{center}
\subfigure[]{\includegraphics[width=0.4\textwidth]{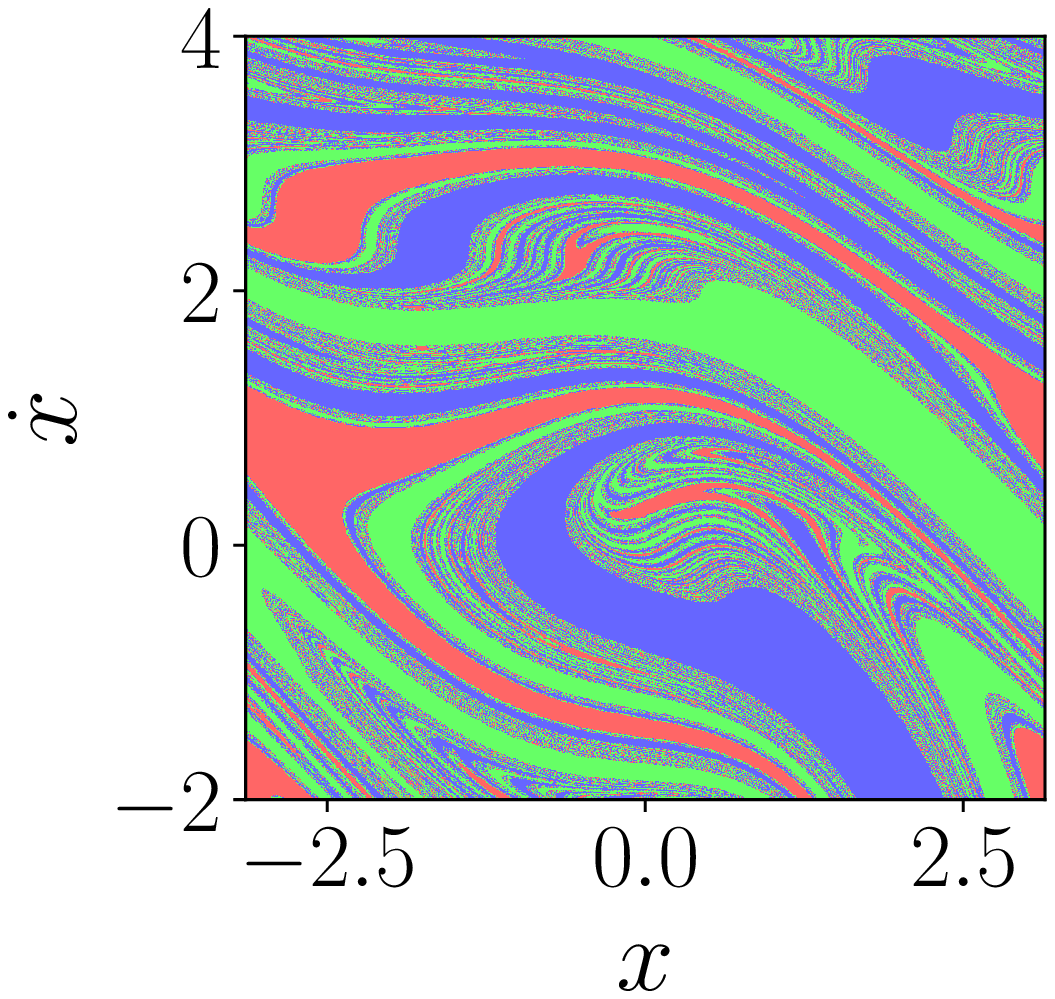}}
\subfigure[]{\includegraphics[width=0.4\textwidth]{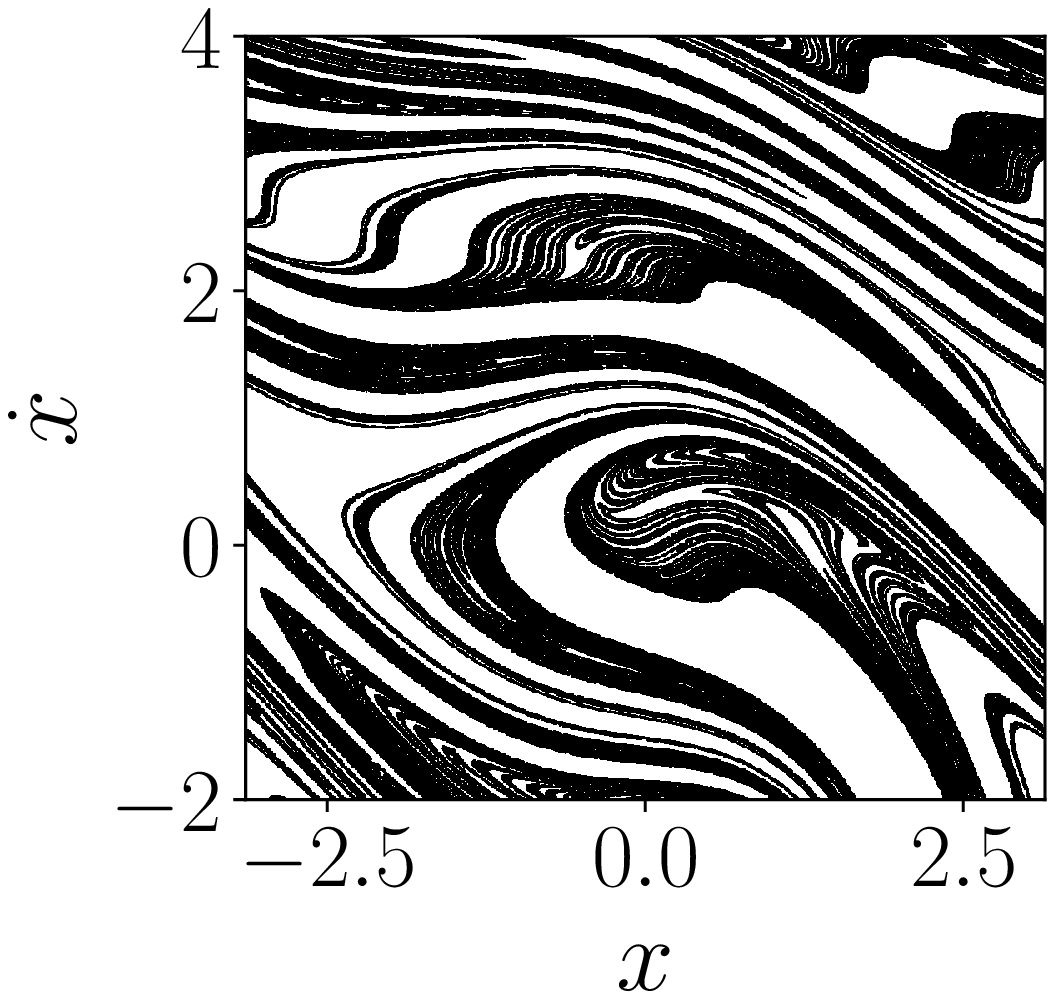}}
\subfigure[]{\includegraphics[width=0.4\textwidth]{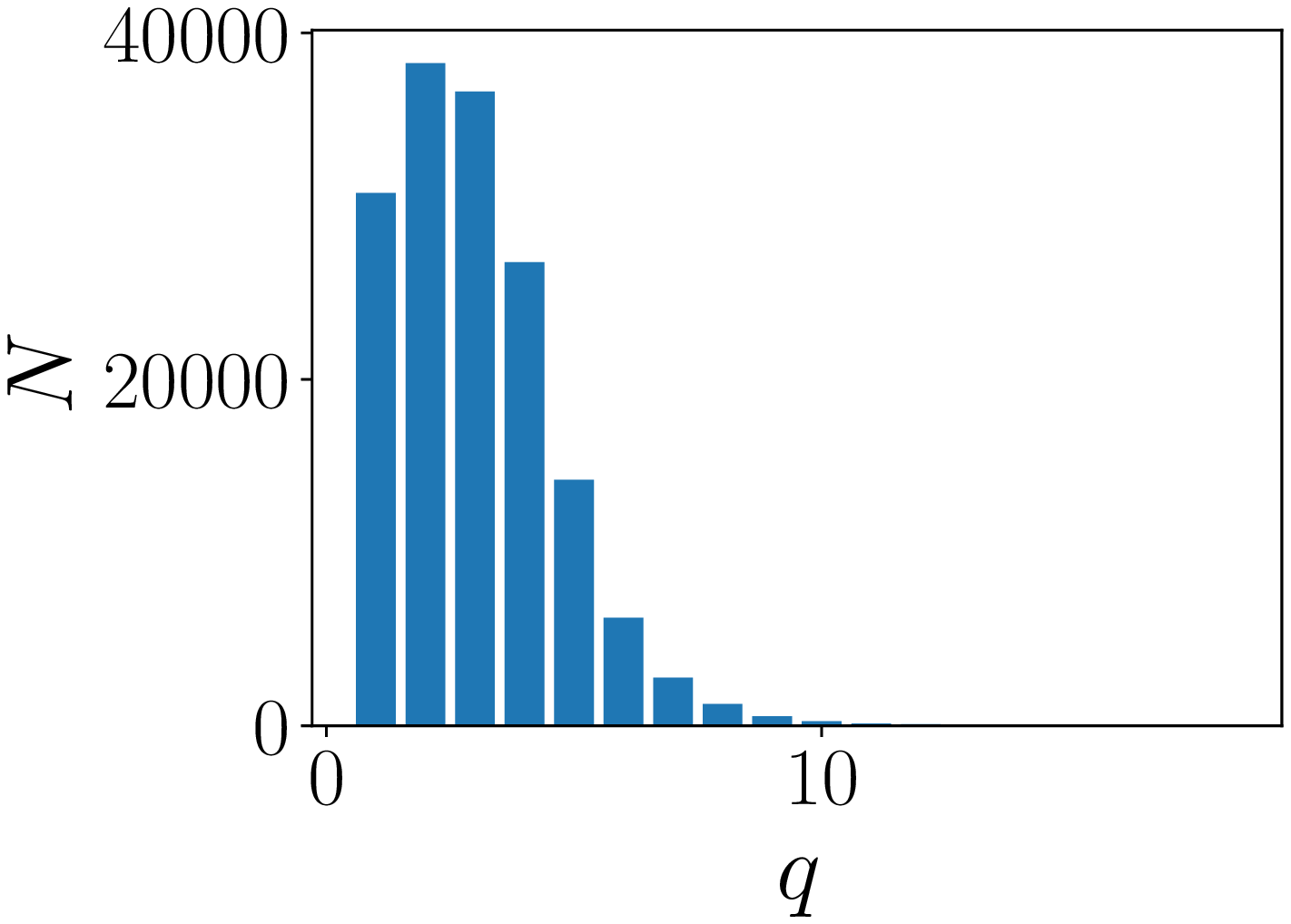}}
\subfigure[]{\includegraphics[width=0.4\textwidth]{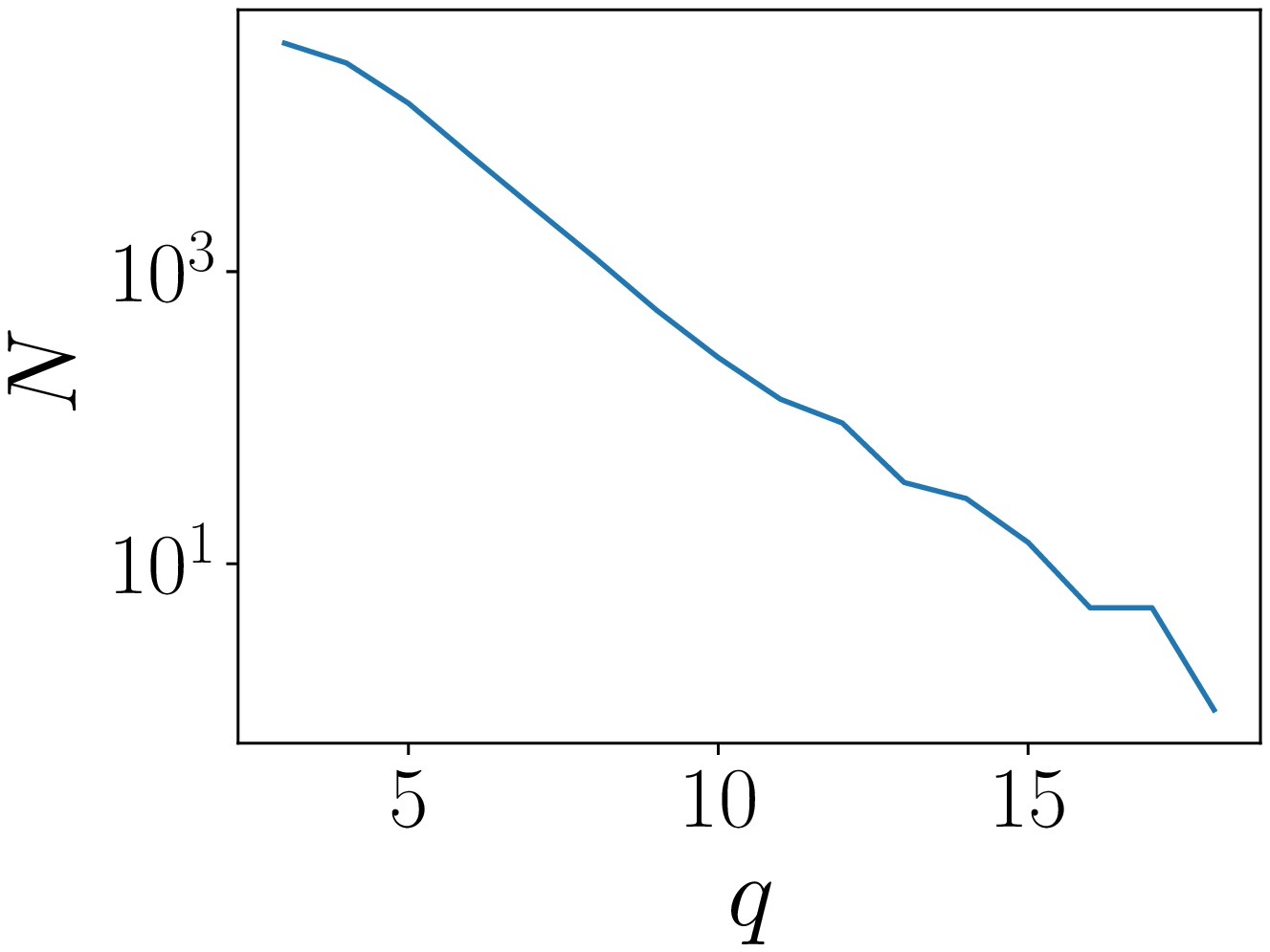}}

\end{center}
\caption{\label{fig_grid_method:wada} {\bf Wada detection with the grid method}. (a) Basin of attraction of the forced damped pendulum $\ddot{x}+0.2\dot{x}+\sin x=1.66\cos t$, (b) All $1000 \times 1000$ boxes are labeled either in the interior (white) or in the boundary of the three basins (black). (c) Histogram showing the number of points $N$ that take $q$ steps to be classified as boundary of three basins. (d) After reaching a maximum, there is an exponential decay of the computational effort related to the fractal structure of the basins. The log-plot reflects this tendency.}
\end{figure}%

When applied to these basins, the grid method classifies all the boxes on the boundary (see Fig.~\ref{fig_grid_method:wada}(b) as Wada after a small number of steps (below $q=18$). The graph of Fig.~\ref{fig_grid_method:wada}(c) shows the decay in the number of boxes that are classified as boundary of three basins. After a peak at $q=3$, the computational effort needed to classify the boxes diminishes. Notice in Fig.~\ref{fig_grid_method:wada}(d) the exponential decay of the number of boxes classified as being in the boundary of three. This decay is related to the fractal structure of the basin. Remarkably, although the number of new trajectories calculated in each stage scales exponentially, the number of boxes that need to be checked decreases exponentially as well, so that the algorithm can be applied in a reasonable time. Indeed, because of this, the performance is better in Wada basins than in partially Wada cases, as we show next.

\begin{figure*}
\begin{center}
  \subfigure[]{\includegraphics[width=0.4\textwidth]{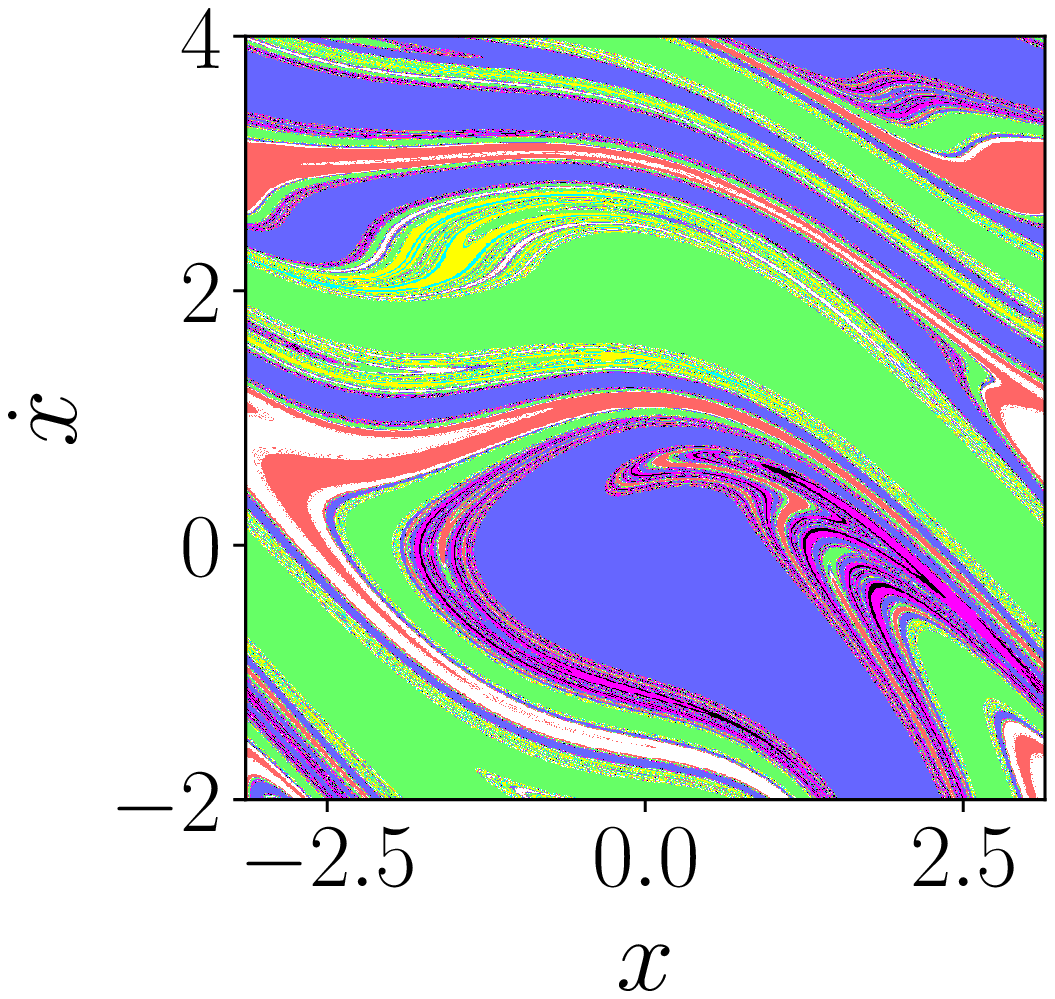}}
  \subfigure[]{\includegraphics[width=0.4\textwidth]{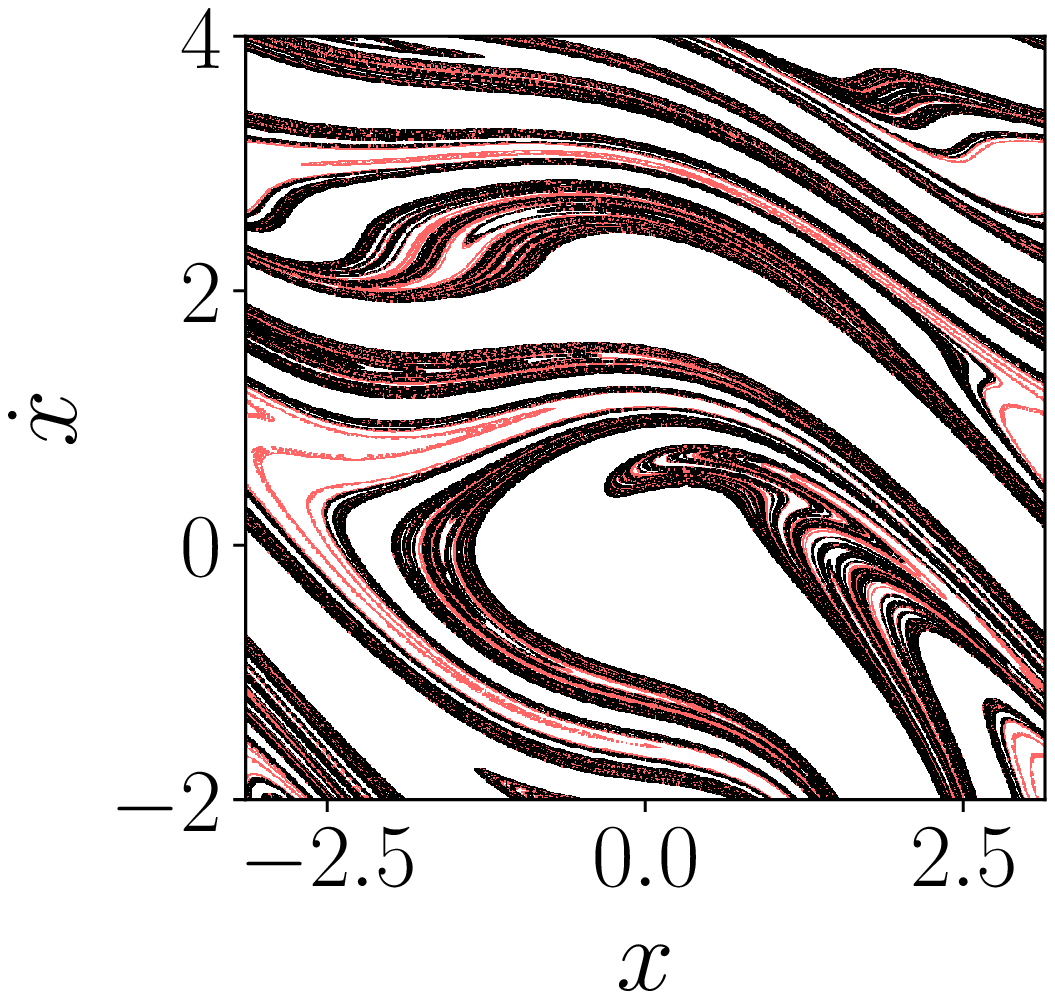}}
  \subfigure[]{\includegraphics[width=0.4\textwidth]{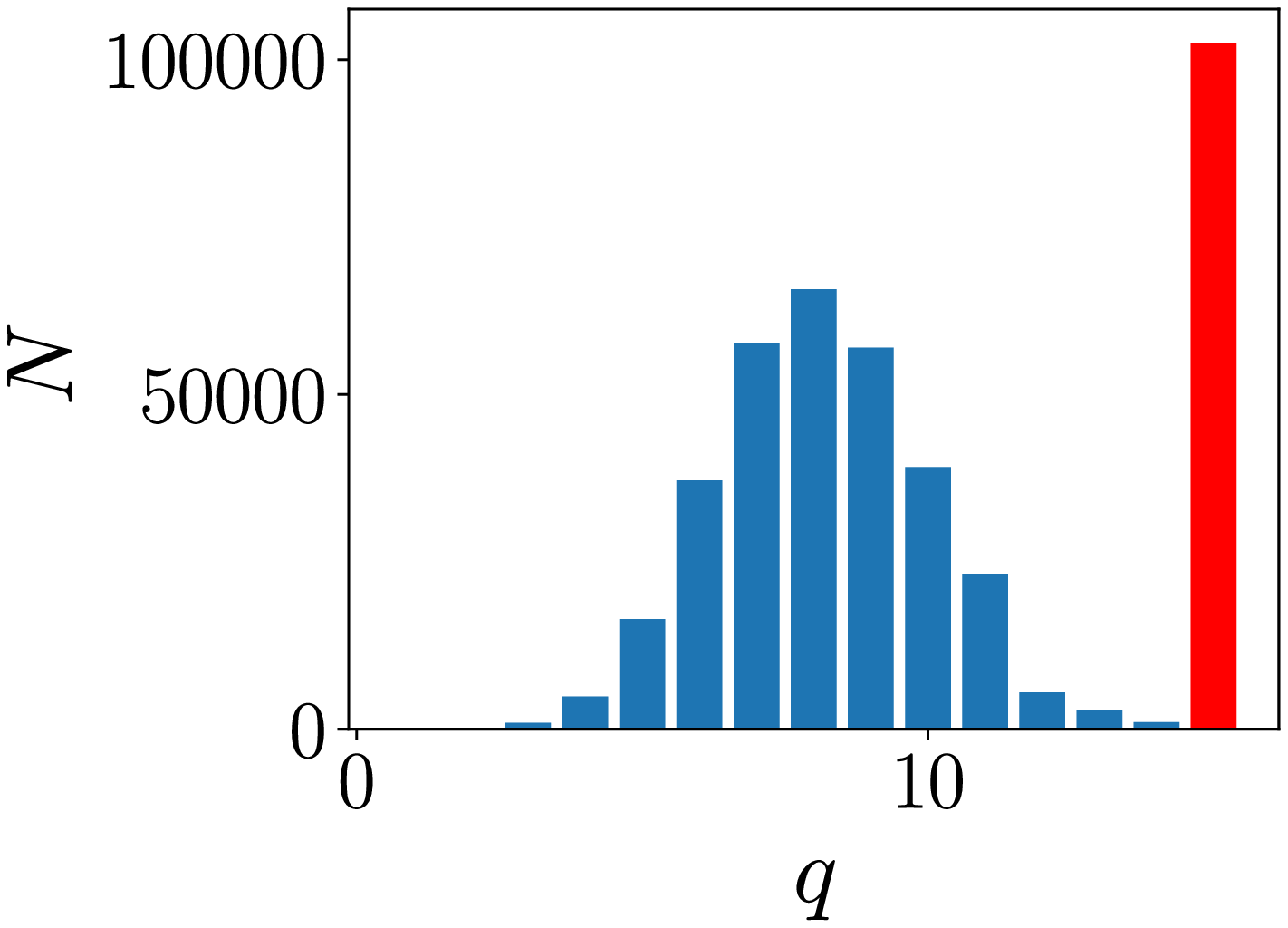}}
  \subfigure[]{\includegraphics[width=0.4\textwidth]{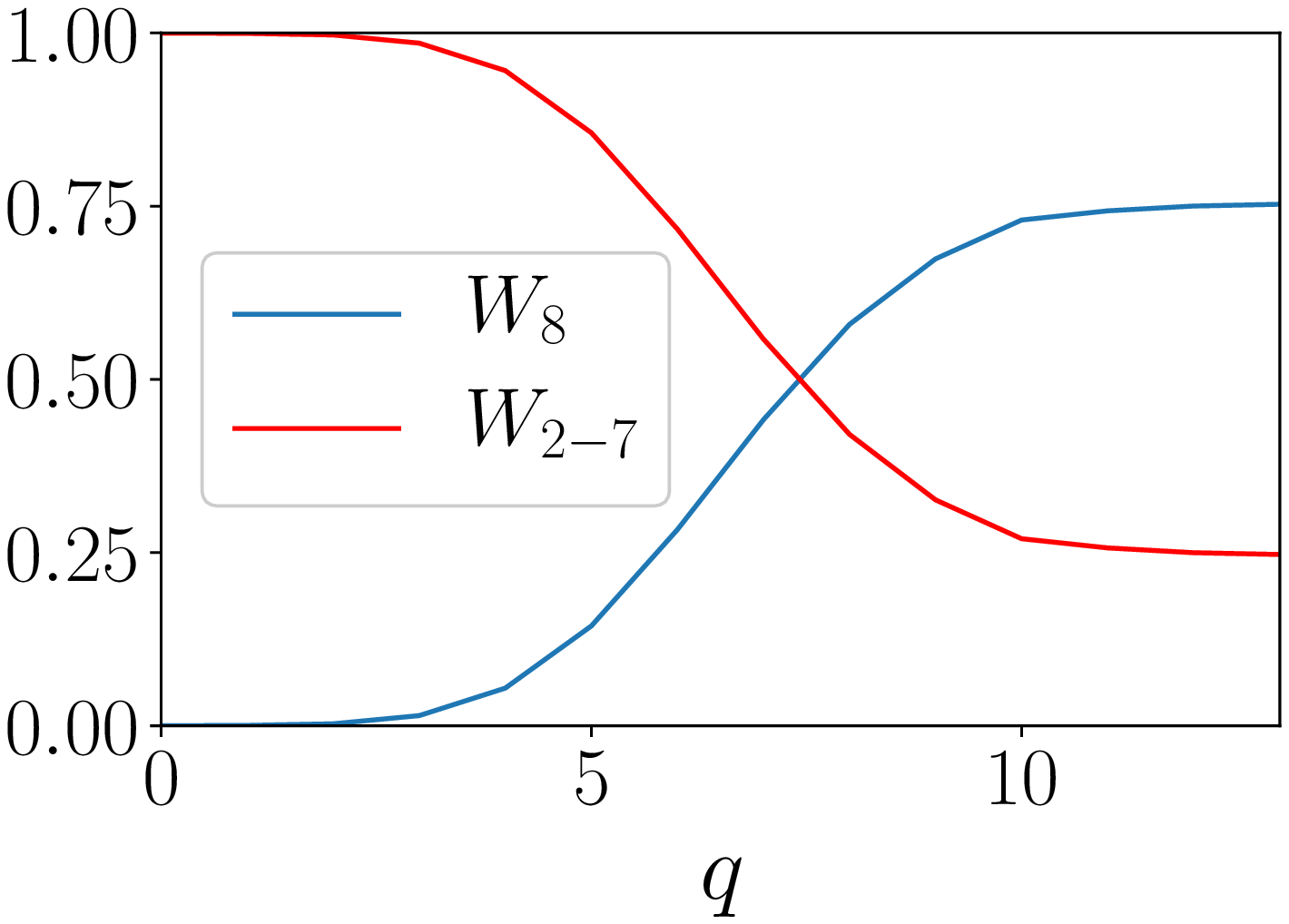}}
\end{center}
\caption{\label{fig_grid_method:partial} {\bf Forced damped pendulum with eight basins}. (a) The following damped forced pendulum $\ddot{x}+0.2\dot{x}+\sin x=1.73\cos t$ shows eight basins of attraction mixed intricately. (b) Some boxes are classified to be in the boundary of eight basins (black dots), but not all of them (red dots), which is a clear example of a partially Wada basin. (c) The computational effort presents the usual shape for the Wada boundary, but the points which are not Wada keep refining until the algorithm meet the stop criterion at $q=15$ (the red bar at rightmost represent the number of boxes not classified as Wada at this stage.). The grid method works best in systems with the Wada property. (d) Evolution of the proportion of boxes in the Wada boundary ($W_8$ in black) and proportion of boxes in a boundary which is not Wada ($W_{2 - 7}$) as a function of the $q$-step. The convergence of $W_8$ is used to determine the stopping rule.\\}
\end{figure*}

The second example is again the forced damped pendulum, but with slightly different parameters given by $\ddot{x}+0.2\dot{x}+\sin x=1.73\cos t$. Now the system has eight basins, depicted in Fig.~\ref{fig_grid_method:partial}(a). The grid method classifies this case as a partially Wada basin after $q=10$ steps. This can be decided when the parameter $W_8$, that gives us the proportion of boxes in the boundary of eight basins, is lower than 1, as seen in Fig. \ref{fig_grid_method:partial}(d). This indicates that not all the boxes on the boundary are in the boundary of the eight basins. Also, the value of $W_8$ can be used as a stopping condition: $W_8$ remains constant after $q \gtrapprox 10$, meaning that no new Wada points are being found in finer resolutions. In this regime, the computational cost increases exponentially in each stage, since the number of new computed trajectories keeps growing, while the number of boxes that are checked remains constant. The red bar in Fig. \ref{fig_grid_method:partial}(c) is the number of boxes that will keep refining indefinitely.

In the original paper, the grid method was illustrated using discrete maps too and, after that, it has been successfully applied to the subspaces of delay differential equations \cite{daza_wada_2017} and open Hamiltonian systems \cite{mathias2017fractal}. It is also important to clarify that given the finite resolution of the grid method, it would classify slim fractals \cite{chen2017slim} as Wada. From a purely mathematical point of view, these boundaries should not be Wada since in the infinity it would not be possible to find the third color. However, from any practical perspective, slim fractals may also display the Wada property at all accessible scales and the grid method is able to correctly account for it.

\section{\label{sec:merging_method}Fusion of colors: the merging method}

In this section, we present the second method to test Wada basins. We call it the \textit{merging method} because it is based on the following observation about Wada basins:
\begin{displayquote}
{\it Wada basins can be merged and their boundary does not change.}
\end{displayquote}

Now let us set some definitions to be rigorous about the precise meaning of the previous statement. We say that a point $p$ is in the \emph{boundary} of a basin $B_i$ if  $\forall\varepsilon>0$, the open ball centered in $p$ of radius $\varepsilon$, $b(p,\varepsilon)$, is such that $b(p,\varepsilon)\cap B_i \neq \emptyset$ and $b(p,\varepsilon)\cap  {B_i}^\complement \neq \emptyset$, where ${B_i}^\complement$ is the complement of $B_i$. If the point satisfies the previous condition for all the basins $B_i$ with $N_A \geq 3$ basins of attraction, we call it a \emph{Wada point}. If all the boundary points are Wada points, then the basin of attraction has the \emph{Wada property}, and we call it a Wada basin.



Assuming that we have $N_A \geq 3$ basins of attraction and each basin $B_i$ has a boundary $\partial B_i$ that we want to determine. A way to identify the points in the boundary $\partial B_i$ is to prove that the point $p$ is arbitrarily close to the set $B_i$ and arbitrarily close to at least one of the other basins $B_j$. That is, $p$ is in the boundary $\partial B_i$ if  $\forall\varepsilon>0$ the open ball centered in $p$ of radius $\varepsilon$, $b(p,\varepsilon)$, is such that $b(p,\varepsilon)\cap B_i \neq \emptyset$ and $b(p,\varepsilon)\cap  \bigcup\limits_{j\neq i} B_j \neq \emptyset$.

With this definition, we can obtain as many different boundaries $\partial B_i$ as possible attractors, since they represent the boundary between the basin $B_i$ and all the other merged basins $\bigcup\limits_{j\neq i} B_j$. We are now ready to provide an alternative (but equivalent) definition of Wada basins: the basins are Wada if and only if the boundaries obtained with the previous procedure are the same, that is $\partial B_i=\partial B_j$ for $\forall i \neq j, i=1,\ldots, N_A$.

This alternative definition emphasizes the fact that two Wada basins can be merged without changing the boundary. More precisely, it is possible to merge up to $N_A-1$ basins without any change in the boundary for $N_A\geq 3$.

\begin{figure*}
\begin{center}
\subfigure[]{\includegraphics[width=0.45\textwidth]{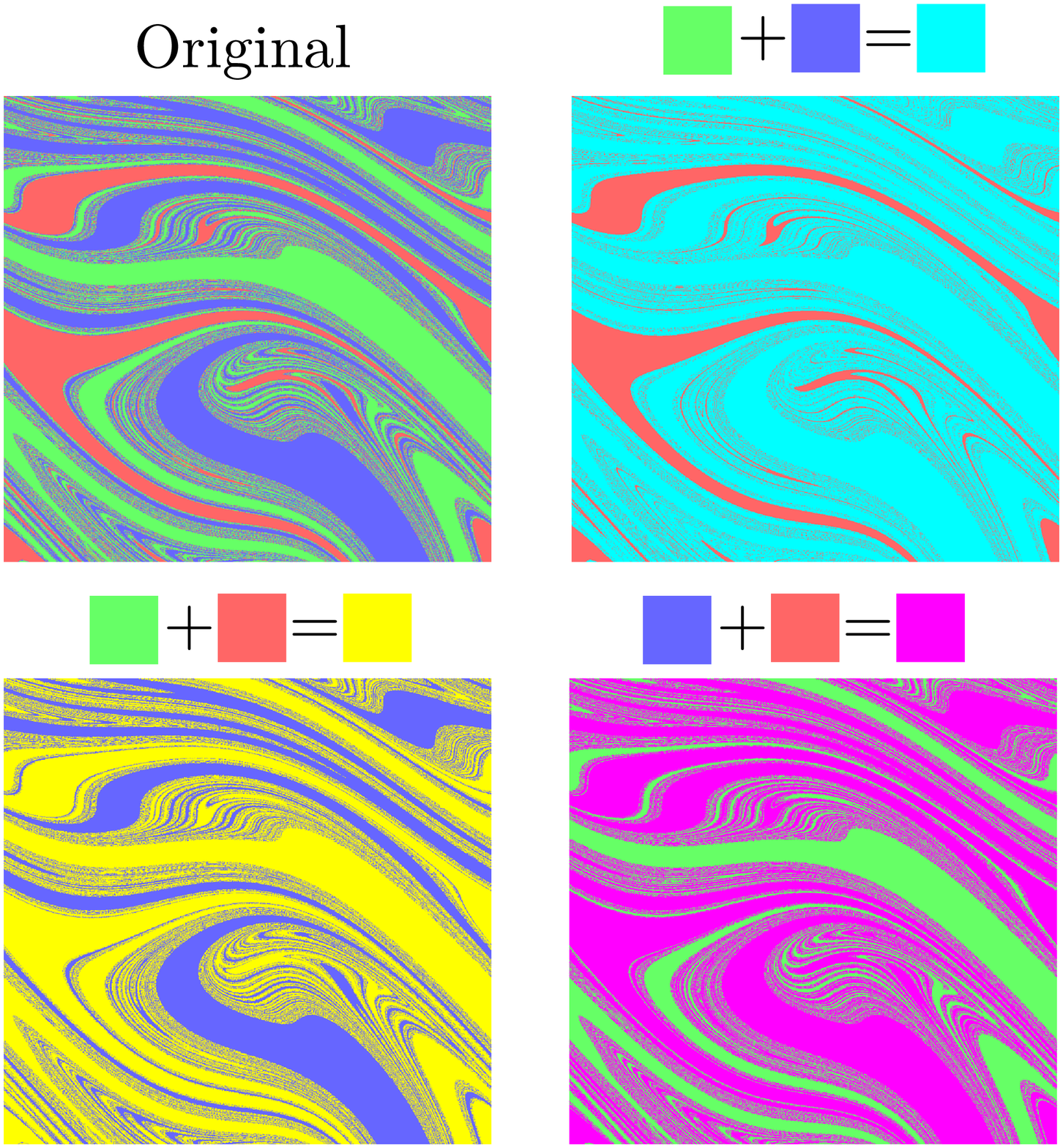}}
\hspace{0.5cm}
\subfigure[]{\includegraphics[width=0.45\textwidth]{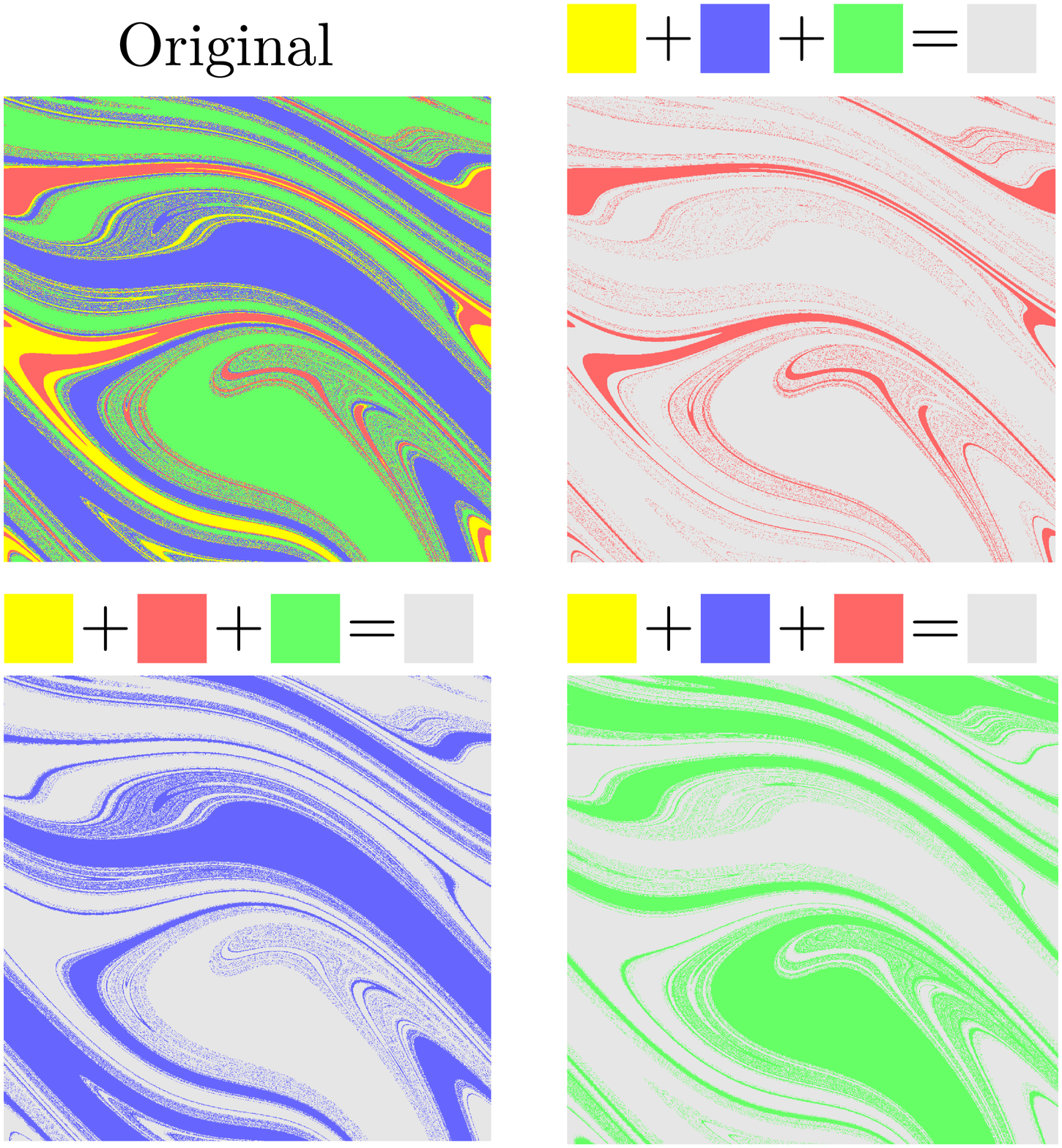}}
\end{center}
\caption{{\bf Graphical description of the merging of basins}. In (a) upper left corner we have the original basin of the forced damped pendulum described by  $\ddot{x}+0.2\dot{x}+\sin x=1.66\cos t$. The other three panels are the modified basins with two merged basins. The colors above indicate which of the original basins have been merged. In (b) The case of the damped pendulum defined by  $\ddot{x}+0.2\dot{x}+\sin x=1.71\cos t$ is shown, which possesses four attractors. We have displayed only three of the four possible combinations of merging. However, these examples are enough to show that the boundaries are not identical.}
\label{merging_figure}
\end{figure*}

As before, we illustrate the merging property using the paradigmatic forced damped pendulum described by Eq.~\ref{forced_dmp_eq}, that is, $\ddot{x}+0.2\dot{x}+\sin x=1.66\cos t$. The upper-left panel of Fig.~\ref{merging_figure}(a) shows the three basins with the Wada property. The other three panels display the basins of attraction that result from the merging of two attractors into one. On top of each basin,, we indicate the colors that have been merged together (yellow=red+green, magenta=blue+red, cyan=blue+green). It is important to notice that each color represents a different basin, being impossible to establish a one-to-one correspondence between basins of different colors. Although the four basins are different, the boundaries are the same in all the cases, as we show numerically in the next section.


We can see how the merging operation works in non-Wada basins. The upper-left panel of Fig.~\ref{merging_figure}(b) shows the basins of the forced damped pendulum defined by $\ddot{x}+0.2\dot{x}+\sin x=1.71\cos t$, which possesses four attractors. In the other three panels, we have merged three basins into a single color gray to improve the contrast of the boundary. If we compare the results of the merging pairwise, we can observe significant differences between boundaries. The aim of the algorithm described in the next section is to quantify these discrepancies numerically.

\subsection{\label{sec:MergingMethod}Description of the merging method}

The property that we have just described, that is, that Wada basins can be merged without any change in their boundary, can be used to build a numerical method to test the Wada property. Formally, all we have to do is to check that the fractal boundaries are the same under the merging of the basins. While it seems an easy task to compare sets visually, it is a very hard problem numerically. This is because in practice, we always have a finite resolution and a restricted set of points.


A usual way to compute the basins of attraction is to select the initial conditions on a grid with linear size $\varepsilon$. The initial condition is at the center of a square pixel of size $\varepsilon$ that we color according to the final state determined by this initial condition. The resolution of the computed boundaries will be limited by the size of this pixel, i.e., by $\varepsilon$. The boundaries computed from merged basins, called the \textit{slim boundaries}, may be slightly different even though we have Wada basins. They are not strictly identical due to the finite resolution imposed by $\varepsilon$, and this holds in spite of any way of computing the basins.

\begin{figure}
\begin{center}
\includegraphics[width=7cm]{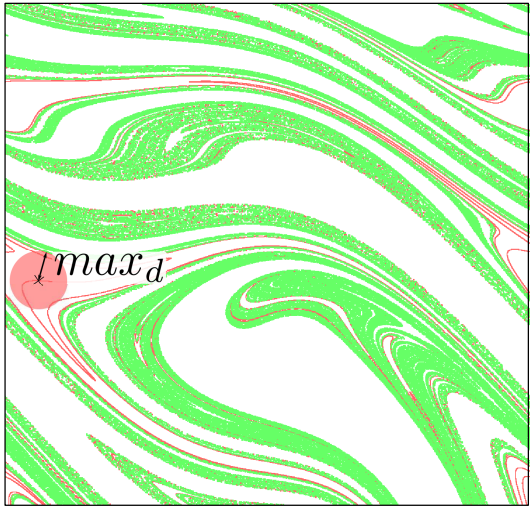}
\end{center}
\caption{\label{pic_hausdorff} \textbf{Interpretation of the Hausdorff distance.} The figure represents two superimposed slim boundaries computed from two different merged basins. One of the boundaries is plotted with red pixels and the other one with green pixels. While it appears that most of the boundaries overlap, some parts of the red boundary do not coincide with the green boundary. The largest distance between the two boundaries is represented by a red circle of radius $max_d$ that corresponds to the Hausdorff distance between the two sets of points.}
\end{figure}

Then the following question arises: how can we compare these boundaries and give a reasonable measure of their similarity? In \cite{daza_ascertaining_2018}, the authors propose to {\it fatten} the slim boundaries replacing each pixel of the boundary by a new fat-pixel of radius $r$. The result is a {\it fat boundary} that looks similar to the original slim boundary but with a thicker stroke. Once all the fat boundaries $\boldsymbol{\partial B_i}$ have been obtained, the algorithm checks whether all the slim boundaries $\partial B_i$ fit in the fat boundaries $\boldsymbol{\partial B_j}$ pairwise such that $\partial B_i \subset \boldsymbol{\partial B_j}$ $\forall i,j=1,\ldots,N_A$. If the test is successful, we say that the basin has the Wada property for the fattening parameter $r$. If the test fails, we can increase the radius $r$ until a radius $r_{max}$ fixed beforehand is reached.

Here, we propose a modification of this technique using the Hausdorff distance \cite{edgar2007measure} that measures the longest possible distance (for a given norm) that we must travel to go from one set to the other set. For a given distance $d_H$ between two sets, we can be sure that any pair of points of the two sets are at a distance $d\leq d_H$.

Mathematically, we must first define the distance between a single point $\bf x$ and the set $Y$:
\begin{equation}
  d({\bf x},Y) =\min_{{\bf y} \in Y}(|| {\bf x} - {\bf y}||),
\end{equation}
so that the Hausdorff distance can be defined as:
\begin{equation}
d_H(X,Y) = \max\{\sup_{{\bf x} \in X} d({\bf x},Y); \sup_{{\bf y} \in Y} d({\bf y},X)\}.
\end{equation}
Computing $d_H$ involves finding the minimum distance for each point of each set. A very large number of pairwise comparisons may be needed if we proceed systematically. Fortunately, there are efficient algorithms to find the nearest neighbors between two large sets of points such as the $k$-$d$ tree algorithm \cite{Friedman_kdtree}. The comparisons can be shrunk down to a matter of seconds in a regular workstation.

Therefore, after merging the basins and obtaining the slim boundaries $\partial B_i$, the next step of the procedure is to measure the Hausdorff distances $d_H(\partial B_i, \partial B_j)$ for each pair of boundaries. We represent an example of distance computed between two different slim boundaries in Fig. \ref{pic_hausdorff}. Among all these distances, it will be useful to know the maximum and minimum values $max_d$ and $min_d$ for further purposes. We can connect this with the definition of a basin with the Wada property at the beginning of the section: the algorithm checks if the points $p_i$ in the boundaries $B_i$ are within a ball $b(p_j,max_d)$ of radius $max_d$ around the points $p_j$ of the boundary $B_j$.

As a simple rule of thumb to quickly check if the system has the Wada property, we can test if $max_d >> min_d$. If this is the case, it is likely that at least two of the boundaries are different. If these two quantities are similar, then a further analysis is needed to decide whether this distance is small or large compared to the size of the phase space under consideration. In any case, it is difficult to give a clear cut and general criterion to decide when a given system possesses the Wada property. However, we will give examples that will illustrate the use of this distance in the next section.

The whole procedure described before can be fully automated and the only input needed is a finite resolution basin. For basins with a resolution of $1000 \times 1000$ and three different attractors, the merging method takes a few seconds to determine whether a basin is Wada running in a regular workstation.

The Haussdorff distance can also be connected with the fattening method of the original paper \cite{daza_ascertaining_2018}. For a grid of size $\varepsilon$ and a Hausdorff distance $d_h$ between the boundary $\partial B_j$ and the partial boundary $\partial B_i$, the ratio $r=d_h/\varepsilon$ is the fattening parameter $r$ needed to cover the entire set $\partial B_i$.

Next we summarize the steps of the merging method:
\begin{enumerate}
\item The input of the algorithm is a picture of the basins at a given resolution $\varepsilon$.
\item For each basin $B_i$, we merge the other basins obtaining two-color basins of attraction made of the original basin $B_i$ and the merged basin $\bigcup\limits_{j\neq i} B_j$. By this process, we get a collection of $N_A$ pictures with only two colors.

\item We compute the slim boundaries of the merged basins $\partial B_i$. In order to do this, we can simply see if a pixel has pixels of different colors around itself. Given the finite resolution of the basins $\varepsilon$, these boundaries may appear slightly different even for Wada basins. For very large basins we can use efficient numerical techniques of edge detection usual in signal processing of images \cite{shapiro2001computer}.

\item The Hausdorff distance $d_H(\partial B_i, \partial B_j)$ is computed for each pair of slim boundaries. We only keep the maximum and minimum distances $max_d$ and $min_d$.

\item If $max_d>>min_d$, we can discard the hypothesis of having a Wada basin. If $max_d \simeq min_d$ and $max_d$ is ``small'', we can conclude that the basin has the Wada property.
\end{enumerate}

\subsection{\label{sec_mg:examples}Examples}

We describe here some results of the detection of Wada and partially Wada basins by means of the merging method. The algorithm is tested for three different systems:
\begin{enumerate}
\item The forced damped pendulum as described in Eq.~\ref{forced_dmp_eq}, $\ddot{x}+0.2\dot{x}+\sin x=F\cos t$ for three different forcing amplitudes $F=1.66$, $F=1.71$, and $F=1.73$. The corresponding basins have, three, four and eight attractors respectively, and only the basin with three attractors has the Wada property.
\item The Hénon-Heiles Hamiltonian \cite{aguirre_wada_2001} described by the equation $H=\frac{1}{2}(\dot{x}^2+\dot{y}^2)+\frac{1}{2}(x^2+y^2)+x^2y-\frac{1}{3}y^3$ and for values of the energy above the critical level $E_c=1/6$ possesses three escape basins in phase space. Here we use two different values of the energy $E>E_c=1/6$, so that we obtain three escape basins that possess the Wada property, though different fractal boundaries.
\item The Newton's method to find complex roots \cite{epureanu_fractal_1998, ziaukas2017fractal}, which is represented by the map $z_{n+1}=z_n-(z^N_A-1)/(r z^{N_A-1})$ with $N_A$ represents the number of basins of attraction.
\end{enumerate}

\begin{table}
\begin{tabular}{|l|c|c|c|c|}
\hline
Dynamical system & $max_d$ & $min_d$ & $(max_d - min_d)/ min_d$ & Wada? \\
\hline
Forced pendulum $N_A =3$ & 0.0365 & 0.0219 & 0.667 & YES \\
\hline
Forced pendulum $N_A =4$ & 0.368 & 0.0439 & 7.3826 & NO \\
\hline
Forced pendulum $N_A =8$ & 0.3976 & 0.0655 & 5.0702 & NO \\
\hline
Hénon-Heiles Hamiltonian $E_0 = 0.2$ & 0.0206 & 0.0168 & 0.2262 & YES \\
\hline
Hénon-Heiles Hamiltonian $E_0 = 0.3$ & 0.0240 & 0.0236 & 0.0169  &  YES \\
\hline
Newton method $N_A=3$ & 0.0300 & 0.0240 & 0.2499  &  YES \\
\hline
Newton method $N_A=4$ & 0.0402 & 0.0350 & 0.1485 &  YES \\
\hline
Newton method $N_A=5$ & 0.0902 & 0.0420 & 1.1476 &  YES \\
\hline
Newton method $N_A=6$ & 0.0780 & 0.0566 & 0.3780 &  YES \\
\hline
\end{tabular}
\caption{{\bf Results of the computation of the Wada merging method for different systems with fractal basin boudaries.} Some of these examples show a fractal basin according to the merging method. All the basins have been computed with a finite resolution of $1000 \times 1000$.}
\label{tab_res_merge}
\end{table}

In Tab.~\ref{tab_res_merge} we summarize the results of the algorithm for the three different systems. In the case of the basins with the Wada property, the relative distance $(max_d -min_d) / min_d$ is usually smaller than 1 (exept for one case). Also, the minimum distance $min_d$ is in all cases two orders of magnitude lower than the size of the phase space, so we can consider this number small and therefore the results accurate.

We can see that in the two examples of fractal basins without the Wada property the ratio $(max_d -min_d) / min_d$ is much higher than the other cases. At any rate, it is up to the user of the method to decide in the end if the basin has the Wada property for this resolution. For a more accurate response, we present in this review  two other numerical methods that may satisfy any need.

\section{\label{Sec.straddle}Find the chaotic set: the saddle-straddle method}

To complete the catalog of numerical methods to detect the Wada property, we present a method that relies on the chaotic dynamics of the system. So far, we have been focused on the structure of the Wada basins. Here we concentrate on a property of these basins directly linked to the dynamics, that is, the existence of an special subset of the boundary, the chaotic saddle, that for Wada basins is the only one existing since there is only one common boundary.

Connected Wada basins are separated by a single connected boundary \cite{kennedy_basins_1991}. In terms of the dynamics, this means that there is a single invariant set under forward iteration, i.e., there is only one stable manifold. As shown by Kuratowski \cite{kuratowski_sur_1924}, this manifold must be an indecomposable continuum. The existence of only one stable manifold involves the existence of a only one saddle. Following these arguments, we can conclude what constitutes the key observation of this third method:
\begin{displayquote}
{\it Connected Wada basins do happen in systems with three or more possible basins and only one saddle, which must be a chaotic saddle.}
\end{displayquote}

Therefore, a numerical proof showing that there is only one chaotic saddle in phase space, would prove the basins to be Wada. We can construct such a proof by combining two different techniques: the merging method as seen in Sec.~\ref{sec:merging_method} and the saddle-straddle algorithm \cite{battelino1988multiple,nusse2012dynamics}. This later algorithm produces a certain number of points arbitrarily close to the chaotic saddle. For this purpose, the algorithm needs two initial conditions in different merged basins and generates a set of points on the boundary between the merged basins. If we are able to show that the set of points is the same for all the pairs of merged basins, we would succeed in proving that there is only one chaotic saddle and consequently that the basin is Wada.

\begin{figure}
\begin{center}
\includegraphics[width=0.6\textwidth]{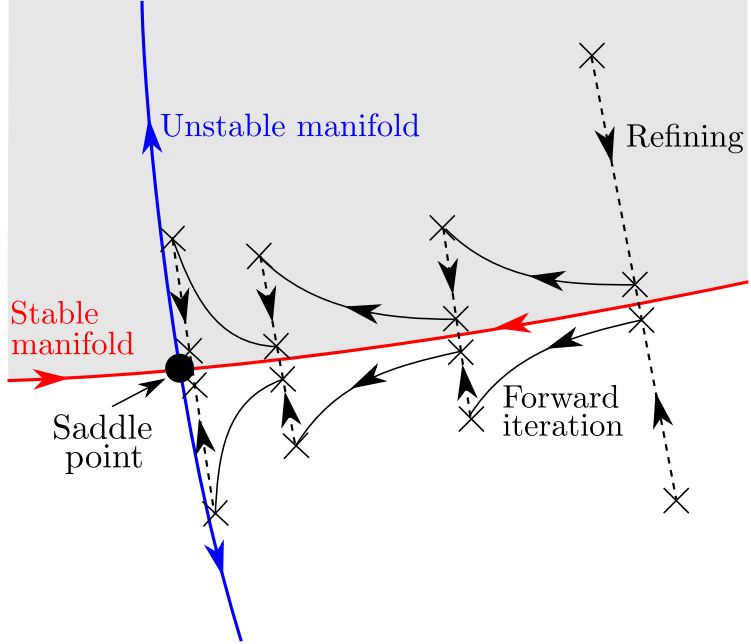}
\end{center}
\caption{\label{fig:ss_method} \textbf{Sketch of the saddle-straddle algorithm.} Initially, two points are selected in such a way that each one lies on a different basin. Then, a bisection method is applied to reduce the distance between the two points to a desired accuracy. After that, the resulting points are iterated and the segment expands, so that the process must start over again. As a result, we obtain a set of arbitrarily small segments straddling the saddle.}
\end{figure}

The saddle-straddle algorithm starts with the two initial conditions on both sides of a boundary in different basins. Using the bisection method, the segment connecting the two initial conditions is shrunk to a very small segment of size just about $10^{-8}$ straddling the boundary. As shown in Fig. \ref{fig:ss_method}, the end points of the segment are iterated forward under the dynamics of the system \footnote{Notice that this implies some sort of time discretization of the system by defining a Poincaré section for example.} and expands naturally due to the vicinity of the unstable manifold, while the stable manifold attracts the segment towards the saddle. As we are pushed away from the boundary, it is necessary to refine again the segment down to a size below $10^{-8}$. The process starts over and we go on with the process until we have reached the saddle after a short transient. After a the desired number of iterations, we have a collection of small segments that are very close to the saddle, we select one endpoint arbitrarily and we end up with an accurate picture of the saddle.

The saddle-straddle algorithm needs two different initial conditions lying in two different merged basins. We must proceed systematically to apply the algorithm to every basin $B_i$ and the basin formed by merging the remainder $\bigcup\limits_{j\neq i} B_j$. In the case that the basins have the Wada property, the chaotic saddles obtained by applying the saddle-straddle algorithm to the different combinations of merged basins must coincide.

In the next section, we give the details of the procedure and we explain how to compare the different sets of points obtained from the algorithm.

\subsection{\label{sec:ss_method}Description of the saddle-straddle method to test for Wada basins}

The saddle-straddle algorithm tracks a saddle that lies in a boundary that separates at least two basins. It is important to know the $N_A$ attractors present in the phase space region that we are analyzing. However, we do not need to compute the basins of attraction, since knowing a set of initial conditions leading to these attractors is enough. We define a pair of basins formed by the basin $B_i$ of the attractor $i$ and the basin $M_i=\bigcup\limits_{j\neq i} B_j$, which is the result of merging the basins of all the other attractors. We obtain $N_A$ different pairs of basins $(B_i, M_i)$ that provide initial conditions for the saddle-straddle algorithm.

In the following, we will use the term \textit{algorithm} to refer to the way of computing the saddles and the term \textit{method} for verifying the Wada property. The algorithm computes a set of segments between basins $B_i$ and $M_i$ arbitrarily close to the saddle at the intersection of the stable and unstable manifolds. The central argument of the method is that if the computed saddles are the same or sufficiently close from each other, then there is only one boundary that separates the $N_A$ basins. In this case, the basins have the Wada property.

As we try to compare the different sets of points representing the saddles, we are faced with a technical challenge. Although the sets are dense in the chaotic saddle, they correspond to different orbits that depend on the initial conditions used for its construction. The chaotic orbits are similar, but they never coincide exactly, making it difficult a direct comparison. However the concept of distance between sets of points is well defined, as already described in Sec.~\ref{sec:MergingMethod}. This distance measures the longest path to connect one set to another set, that is, the largest distance between any two points of both sets.

After solving the problem of comparing chaotic sets, another question arises: when do we consider that two sets belong to the same saddle? What is a small distance between two sets? To answer these questions, we must first define the diameter of a set
\begin{equation}
d_s(A) = \sup\{ ||{\bf x} - {\bf y}  || :  {\bf x},{\bf y} \in A \}.
\end{equation}
To put it simply, it is the largest Euclidean distance between any two points of a set A. If the set is an orbit that belongs to an attractor we have an estimation of the size of the attractor. This allows us to define the following criterion: {\it if the measured Hausdorff distance between the sets is small with respect to the diameter $d_s$ of one of the set, we can say that the sets correspond to the same saddle.}

We can summarize the steps of the method as follows:
\begin{enumerate}
  \item First, we classify the attractors of the dynamical system and we assign an integer $i$ to each basin.
  \item We form the pairs of basins as follows: for each attractor, we define the basin $B_i$ of the attractor and the basin $M_i$ as the union of the remaining basins. There are as many pairs of basins as attractors.
  \item We compute the saddle for each pair of basins using the saddle-straddle algorithm.
  \item The saddles are compared pairwise using the Hausdorff distance $d_H$. We consider that the saddles belong to the same set when the distance $d_H$ is small compared to the diameter of the set $d_s$. In case the saddles have different diameters, we will pick the largest.
  \item If all the previous comparisons are successful, then there is only one boundary and the basins of attraction possess the Wada property.
\end{enumerate}

Notice that if the distance between two sets is of the same order of magnitude as the diameter of the set $d_s$, we can discard the hypothesis of having the Wada property. Another common situation where we can discard the case of Wada basins is when the diameter of the set is very small (about the size of the straddle segment). This is an indication of a saddle point on a smooth boundary that separates two basins.

To correctly measure the distance between the sets, the number of points $n_p$ should be large enough. If the sets do not have enough points the distance $d_H$ might be biased.

\subsection{\label{sec:ss_examples}Examples}

\begin{figure}
  \subfigure[]{\includegraphics[width=6cm]{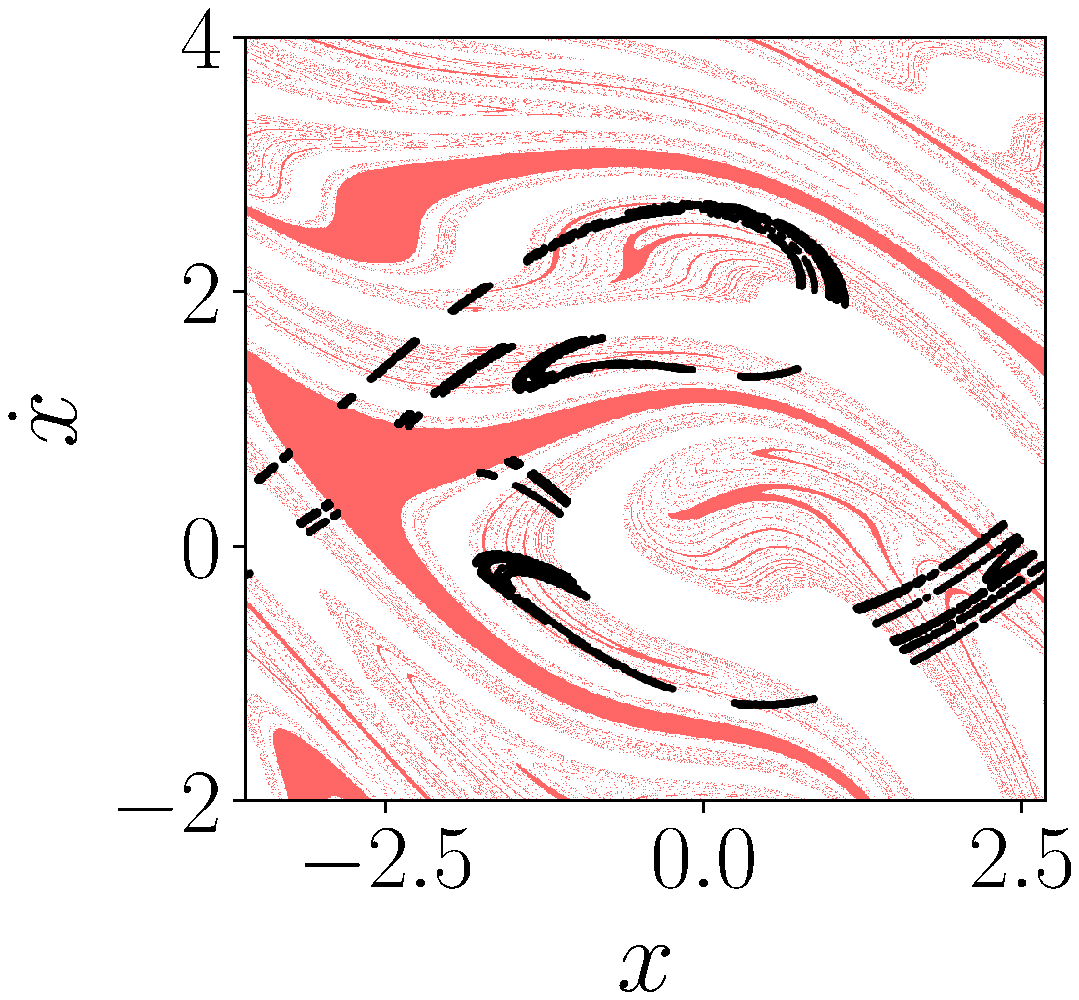}}
  \subfigure[]{\includegraphics[width=6cm]{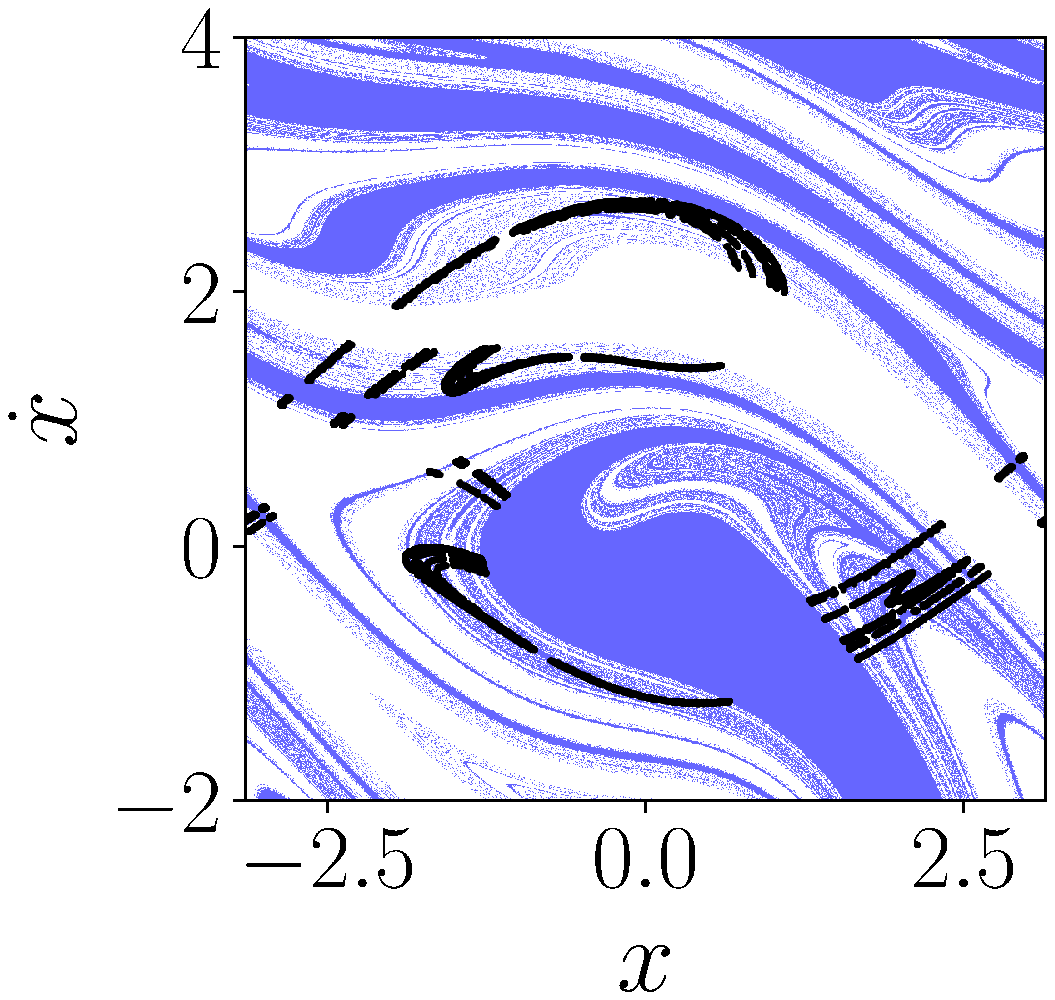}}
  \subfigure[]{\includegraphics[width=6cm]{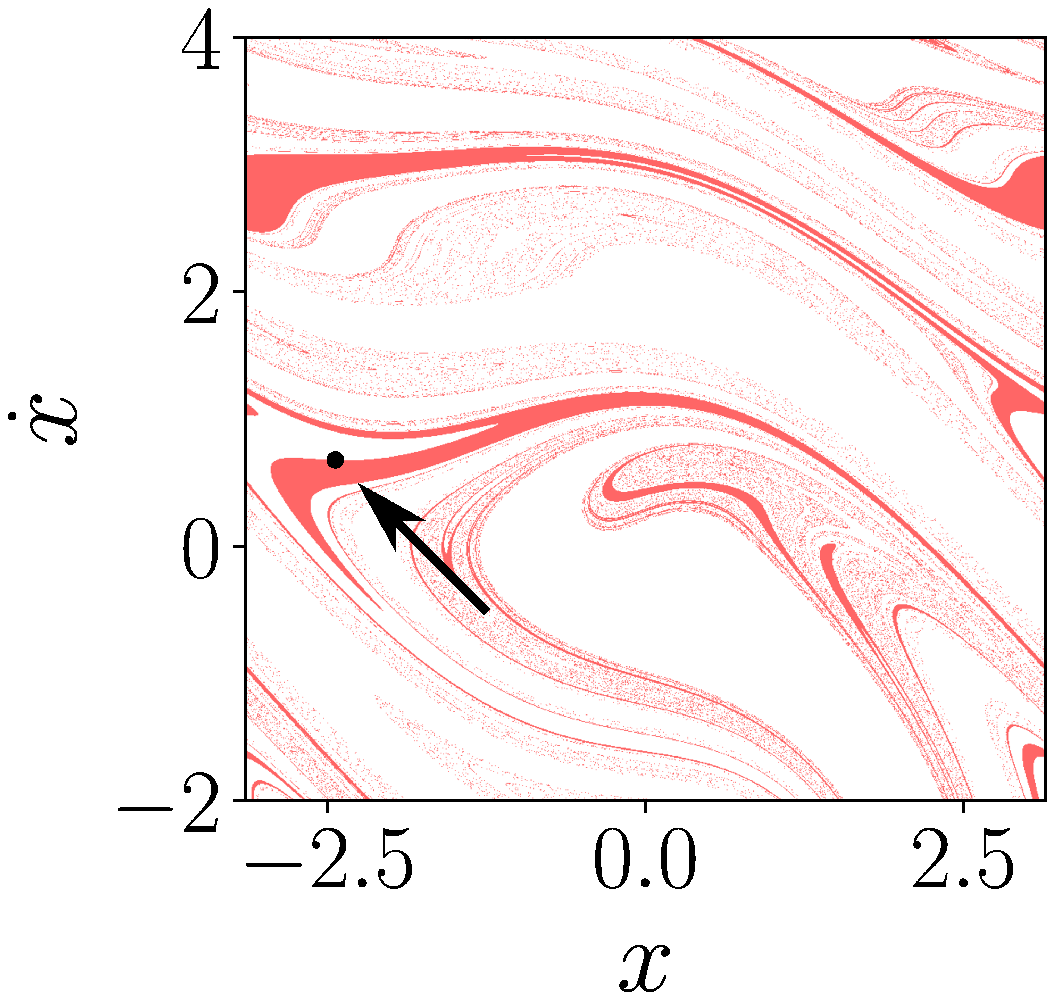}}
  \subfigure[]{\includegraphics[width=6cm]{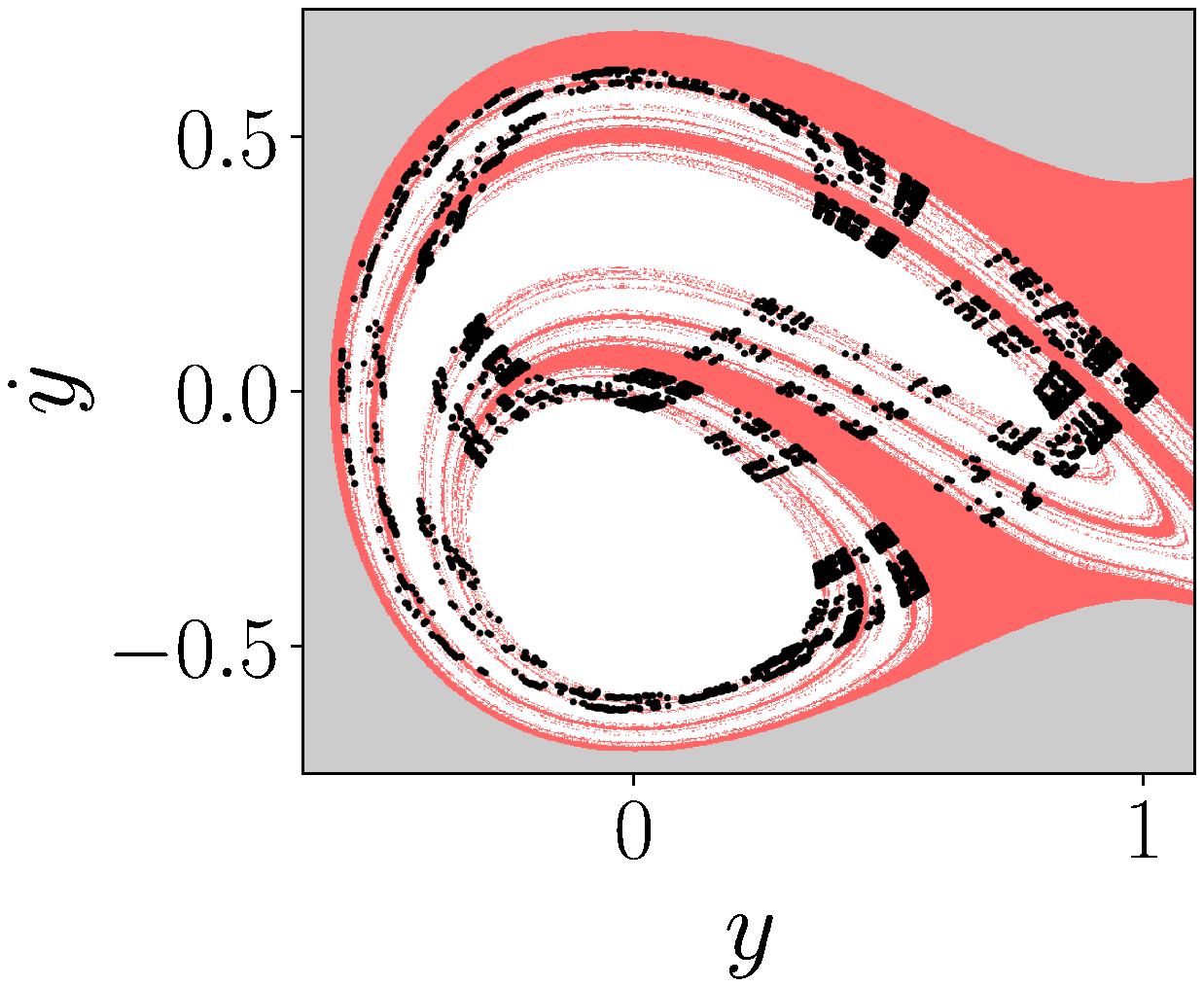}}
  \caption{\textbf{Computations of saddles with the saddle-straddle algorithm.} (a) The picture represents the chaotic saddle embedded in the only boundary of the forced damped pendulum with equation $\ddot{x}+0.2\dot{x}+\sin x=1.66\cos t$. (b) We have represented the computation of the saddle associated to the boundary between basins $B_2$ and $M_2$ of the forced damped pendulum with equation: $\ddot{x}+0.2\dot{x}+\sin x=1.71\cos t$. In (c) we have the saddle corresponding to the boundary between basins $B_1$ and $M_1$. (d) shows the chaotic saddle of the boundary in the Hénon-Heiles Hamiltonian for the energy $E=0.25$.}
  \label{fig.ss_saddles}
\end{figure}

Again, we will test the algorithm on two systems with the Wada and partial Wada property, the forced damped pendulum and the Hénon Heiles potential. As we have shown earlier, the forced damped pendulum with three atractors shows the Wada property. In Fig.~\ref{fig.ss_saddles}(a), we show the saddle obtained from the application of the saddle-straddle algorithm to a basin $B_1$ and a merged basin $M_1$ of this system. We can see that the saddle is embedded in the fractal boundary between them. There is only one saddle as it can be interpreted from the results of the Hausdorff measure between chaotic sets. We denote by $S_i$ the saddle obtained from the pair of basins ($B_i$,$M_i$). The results of the comparisons for 40000 points are: $d_H(S_1,S_2) = 0.04686$, $d_H(S_1,S_3) = 0.04689$ and $d_H(S_2,S_3) = 0.04650$. The distances $d_H$ are very small compared to the diameter of the saddle under study measured as $d_s(S_1)\simeq 2\pi$, which confirms our first impression that all sets of points belong to the same saddle.

In Fig.~\ref{fig.ss_saddles}(b) and (c) we have the case of the partially Wada basin for the forced damped pendulum with three attractors, described by $\ddot{x}+0.2\dot{x}+\sin x=1.71\cos t$. From the two plots we can already conclude that the system is not Wada since one of the saddles is a saddle point on a smooth boundary.The Hausdorff distances computed between each pair of sets for $40000$ points show clearly that there is not only one boundary: $d_H(S_1,S_2) = 5.604$, $d_H(S_1,S_3) = 5.604$, $d_H(S_2,S_4) = 5.604$, $d_H(S_3,S_4) = 5.604$, $d_H(S_1,S_4) = 5.02\cdot 10^{-9} $ and $d_H(S_2,S_3) = 0.064$.
The very small distance $d_H(S_1,S_4)$ between the saddles $S_1$ and $S_4$ shows that the two saddles are identical. Also the diameter of these sets $d_s(S_1)= d_s(S_4) \leq 1\cdot 10^{-8} $ shows without any doubt that there is a single saddle point on a smooth boundary between basins $B_1$ and $B_4$. The algorithm reveals that there is another saddle in the phase space as shown by the distance $d_H(S_2,S_3)$ and diameters $d_s(S_2)= d_s(S_3) = 2\pi$. There are two different saddles and all we can say is that the system has at best the partial Wada property.

Our last example with the Wada property is the Hénon-Heiles Hamiltonian with an energy above the critical value $E=0.25>E_c$ that presents three escape basins. The straddle set $S_1$ obtained from the pair ($B_1$,$M_1$) is shown in Fig.~\ref{fig.ss_saddles}(d). The computation of the Hausdorff distance for $n_p=10000$ gives the following results $d_H(S_1,S_2) = 0.087$, $d_H(S_1,S_3) = 0.058$ and $d_H(S_2,S_3) = 0.085$. Despite the Hénon-Heiles does not have any attractor, it is possible to compare these numbers against the diameter of the saddle $S_1$: $d_s(S_1)=1.5$. The escape basin of this Hamiltonian system has the Wada property according to our procedure: all the distances are small compared to the diameter of the set.

\section{Comparison of available methods to assert the Wada property}

\begin{table*}
\begin{tabular}{|l|p{3cm}|l|c|p{6cm}|}
\hline
\multicolumn{1}{|c}{Name} & \multicolumn{1}{|c|}{Type of system} & Dim. & Computation& \multicolumn{1}{c|}{What we need} \\
    &                 &       & time &  \\

\hline
Nusse-Yorke method \cite{nusse_wada_1996} & ODEs Hamiltonians Maps & 2D & 1$^*$ & It requires a detailed knowledge of the basin and the boundaries (accessible unstable periodic orbit embedded in the basin boundary). \\
\hline
Grid method \cite{daza_testing_2015} & Any dynamical system & n-D & 100  & It requires the basins and the dynamical system to compute parts of the basin at a higher resolution. \\
\hline
Merging method \cite{daza_ascertaining_2018} & Any dynamical system & n-D & 0.01 & It needs to know the basins, but not the dynamical system.\\
\hline
Saddle-straddle method \cite{wagemakers2020saddle} & ODEs Hamiltonians Maps &  2D & 1 & It needs to know the dynamical system, but not the basins.\\
\hline
\end{tabular}
\caption{\label{tab1} Comparison of the principal procedures to test if a basin of attraction has the Wada property. The time noted with $^*$ refers only to the computation time and does not take into account the previous study of the system.}
\end{table*}

We have reported here three different techniques to detect the Wada property in basins of attraction, besides the already known Nusse-Yorke method, each one with its own advantages and drawbacks. Table~\ref{tab1} can serve as a quick guide to pick the right method depending on the nature of the problem. The computation times displayed in Tab.~\ref{tab1} are estimates relative to the the time taken by the saddle stradle method to detect the Wada property of the forced pendulum presented in the previous sections. This task would take about one hour on a normal workstation. Notice that these times may vary depending on the problem, the specific hardware and so on. The effort needed to apply the methods is difficult to evaluate directly because of the required input, such as the basin of attraction on a regular grid.

In the following, we discuss the strengths and weaknesses and we give some indications about the expected accuracy of each method:
\begin{itemize}
  \item {\bf Nusse-Yorke method.} When there is an accessible unstable periodic orbit in the basin boundary that can be tracked, the Nusse-Yorke method is a good candidate. It provides a precise answer to the problem. The problems with the method are related to the need of a detailed study of the dynamical system. In fact, many works have been devoted to the application of this method to just one dynamical system with fixed parameters \cite{poon_wada_1996, toroczkai_wada_1997, aguirre_wada_2001, aguirre_unpredictable_2002}. The computation of the unstable manifold of the periodic orbit can be burdensome in some cases \cite{daza_wada_2017, BHs}. Also the method is restricted to ODEs, Hamiltonians and maps showing the connected Wada property. If the result of the test is a Wada basin, then we have an exact answer as long as we are sure that we have found all the unstable periodic orbit. It is the weak point of the method.

  \item {\bf Grid method.} The grid method is based on the idea that, when a basin has the Wada property, between two initial condition belonging to different attractors we will always find an initial condition leading to a third attractor. It is an interesting method when the basins can be computed easily. It gives a reliable answer with useful information about the structure of the basin. Also, it can be easily automated. However, it can be very slow given that for some boxes the algorithm needs to refine the grid to very small resolutions.

This method can be considered accurate since we have a stopping criterion based on the number $W_m$ that tells us how many boxes are on the boundary of $m$ basins. The algorithm stops when $\vert W_m(step+1)-W_m(step) \vert <\varepsilon$, being $\varepsilon$ a small positive number previously fixed. For the examples presented in the text $\varepsilon=0.005$. It guarantees that the boxes have been correctly classified in the boundary of $N_A$ basins in the case of Wada.

  \item {\bf Merging method.} The merging method to detect the Wada property in basins of attraction hinges on the invariance of the boundary through the merging operation of basins. This is beyond all doubt the fastest method of them all, it is fast and easy to implement (about one hundred lines of code for everything). Also once the basins have been obtained, the method does not assume anything on the underlying dynamics. If the basin is available or can be computed quickly it may be the first method to try. It allows a quick classification of the basins. However, the method is reliable up to the resolution of the basins and spurious or noisy points in the basins can perturb the results of the Hausdorff distances.

 This method gives us two numbers after its application: $min_d$ and $max_d$, the smallest and largest Hausdorff distance between the computed slim boundaries. The researcher must decide with these two numbers in hand whether the basin is Wada or not. A rule of thumb to help this decision is taking the relative distance $(max_d-min_d)/min_d = 2$ as a decision threshold. Above this number we can assume that the system is Wada.

  \item {\bf Saddle-straddle method.} This method is based on a very basic observation: if the basin has the Wada property, then there is only one saddle, and it is chaotic. When the problem is in the plane it is a powerful technique to identify the Wada property. The basins of attraction are not needed since the algorithm relies on the dynamics of the system. We must say that this method is limited to ODEs, Hamiltonians and maps and is unable to detect disconnected Wada boundaries.

The accuracy of the test depends directly on the length of the computed time series, because when two saddles are compared, the Hausdorff distance decreases as a power of the number of points. From our simulations, at least $10^4$ points are necessary to have a Hausdorff distance below $0.01$ for two time-series from the same chaotic saddle. If the Hausdorff distance is two orders of magnitude inferior to the diameter of the set, then the time series belong to the same saddle.

\end{itemize}

We have so far exposed the published techniques available to detect the Wada property. However, other approaches might be possible, as for example a modified version of the saddle-straddle method by using the stable manifold of the saddle instead of the very saddle. The strategies exposed in this article explore the Wada property under different angles, nonetheless we are confident that there might be other ways to tackle this problem.

\section{Conclusions}

Proving the Wada property in dynamical systems may require different approaches adapted to the particularities of the problem under study. We have described several numerical techniques that reflect the state of the art for the study of how to detect Wada basins. One of the fascinating aspects of these techniques is that they all rely on different characteristics that define the Wada property, and which reveal different aspects of this intricate structure.

A possible and important extension is the application of these techniques to higher dimensions. The plane is for sure an important case in the study of dynamical systems, however systems in higher dimension may also present the Wada property. We should discuss the applicability of each method for dimension three and beyond. The grid method would need very little adaptation, the basic principle of ``finding the third color'' is independent of the dimension. Given that the grid method operates within a line between points in two different basins, the dimension of the basin would only affect the performance in the initial computation of the basins, but not in the successive refinements. We could even replace the full computation of the basins by computing a few scattered points in a Monte Carlo fashion and apply the grid method on them. The merging method can also be extended to higher dimensions. The slim boundaries would be sets of dimension $N-1$ that can obtained reasonably fast with filtering techniques. The Hausdorff metric can be computed to compare each sets. Again, the major challenge with this method is the computation of the basin beforehand as the number of grid points grows with a power of $N$. The straddle method is restricted to the plane.

Among possible application of this tool in dynamical system is the study of the space of parameters. The information of the parameter region where the basins are Wada combined with other measures such as the Basin Entropy \cite{daza_basin_2016} and the uncertainty exponent. This would break down the information about the uncertainty of the phase space into three component.

Also, these new methods broaden the scope of application of the original idea of Yoneyama to unexpected fields \cite{BHs}, illustrating that it constitutes a very special property of chaotic dynamical systems usually indicating a lack of predictability and with a bright future ahead in spite of all the work that has been done so far.

\section*{Acknowledgements}

This work has been financially supported by the Spanish State Research Agency (AEI) and the European Regional Development Fund (ERDF, EU) under Projects No.~FIS2016-76883-P and No.~PID2019-105554GB-I00. AD acknowledeges financial support from NSF CIQM Grants No. DMR-12313 19, and NSF CHE 1800101, and the Real Colegio Complutense (RCC) that supported his research at Harvard University.


\end{document}